\documentclass{article}
\usepackage[paperwidth=7.56in,paperheight=10.31in,left=0.69in,right=0.5in]{geometry}

\usepackage{authblk}
\usepackage{comment}
\usepackage{times}
\usepackage{helvet}
\usepackage{courier}
\usepackage{graphicx}
\usepackage{multirow}  
\usepackage{subcaption}
\usepackage{caption}
\usepackage{url}

\usepackage{color,soulutf8}

\usepackage{apacite}

\usepackage{silence}
\usepackage{amsmath}
\usepackage{algorithm}
\usepackage[noend]{algpseudocode}
\makeatletter
\def\BState{\State\hskip-\ALG@thistlm}
\makeatother

\algnewcommand\algorithmicforeach{\textbf{for each}}
\algdef{S}[FOR]{ForEach}[1]{\algorithmicforeach\ #1\ \algorithmicdo}

\usepackage{xcolor}

\setstcolor{red} 

\usepackage[textsize=scriptsize,backgroundcolor=green]{todonotes}
\title{Revisiting Narrow-and-Deep Judging With An Intelligent Topic Selection Method For Information Retrieval Test Collections}
\title{Intelligent Topic Selection for Low-Cost Information Retrieval Evaluation: A New Perspective on Deep vs.\ Shallow Judging}

\begin{document}

\author[1]{Mucahid Kutlu}
\author[1]{Tamer Elsayed}
\author[2]{Matthew Lease}
\affil[1]{Dept.\ of Computer Science and Engineering, Qatar University, Qatar}
\affil[2]{School of Information, University of Texas at Austin, USA}

\maketitle

\begin{abstract}

While {\em test collections} provide the cornerstone for Cranfield-based evaluation of information retrieval (IR) systems, it has become practically infeasible to rely on traditional {\em pooling} techniques to construct test collections at the scale of today's massive document collections (e.g., ClueWeb12's 700M+ Webpages). This has motivated a flurry of studies proposing more cost-effective yet reliable IR evaluation methods. In this paper, we propose a new {\em intelligent topic selection} method which reduces the number of search topics (and thereby costly human relevance judgments) needed for reliable IR evaluation. To rigorously assess our method, we integrate previously disparate lines of research on intelligent topic selection and \emph{deep} vs.\ \emph{shallow} judging (i.e., whether it is more cost-effective to collect many relevance judgments for a few topics or a few judgments for many topics). While prior work on intelligent topic selection has never been evaluated against shallow judging baselines, prior work on deep vs.\ shallow judging has largely argued for shallowed judging, but assuming random topic selection. We argue that for evaluating any  topic selection method, ultimately one must ask whether it is actually useful to select topics, or should one simply perform shallow judging over many topics? In seeking a rigorous answer to this over-arching question, we conduct a comprehensive investigation over a set of relevant factors never previously studied together: 1) method of topic selection; 2) the effect of topic familiarity on human judging speed; and 3) how different topic generation processes (requiring varying human effort) impact (i) budget utilization and (ii) the resultant quality of judgments. Experiments on NIST TREC Robust 2003 and Robust 2004 test collections show that not only can we  reliably evaluate IR systems with fewer topics, but also that: 1) when topics are intelligently selected, deep judging is often more cost-effective than shallow judging in  evaluation reliability; and 2) topic familiarity and topic generation costs greatly impact the evaluation cost vs.\ reliability trade-off. Our findings challenge conventional wisdom in showing that deep judging is often preferable to shallow judging when topics are selected intelligently. 






\end{abstract}

\section{Introduction} 
\label{introduction}
 
Test collections provide the cornerstone for system-based evaluation of information retrieval (IR) algorithms in the Cranfield paradigm~\cite{cleverdon1959evaluation}. A test collection consists of: 1) a {\em collection} of documents to be searched; 2) a set of pre-defined user {\em search topics} (i.e., a set of topics for which some users would like to search for relevant information, along with a concise articulation of each topic as a {\em search query} suitable for input to an IR system); and 3) a set of human {\em relevance judgments} indicating the relevance of collection documents to each search topic. Such a test collection allows empirical A/B testing of new search algorithms and community benchmarking, thus enabling continuing advancement in the development of more effective search algorithms. Because 
exhaustive judging of all documents in any realistic document collection is cost-prohibitive, traditionally the top-ranked documents from many systems are {\em pooled}, and only these top-ranked documents are judged. Assuming the {\em pool depth} is sufficiently large, the reliability of incomplete judging by pooling is well-established \cite{sanderson2010test}.

However, if insufficient documents are judged, evaluation findings could be compromised, e.g., by erroneously assuming unjudged documents are not relevant when many actually are relevant~\cite{Voorhees06}. The great problem today is that: 1) today's document collections are increasingly massive and ever-larger; and 2) 
realistic evaluation of search algorithms requires testing them at the scale of document collections to be searched in practice, 
so that evaluation findings in the lab carry-over to practical use. Unfortunately, larger collections naturally tend to contain many more relevant (and seemingly-relevant) documents, meaning human {\em relevance assessors} are needed to judge the relevance of ever-more documents for each search topic. As a result, evaluation costs have quickly become cost prohibitive with traditional pooling techniques \cite{sanderson2010test}.
Consequently, {\em a key open challenge in IR is to devise new evaluation techniques to reduce evaluation cost while preserving evaluation reliability}. In other words, how can we best spend a limited IR evaluation budget?

A number of studies have investigated whether it is better to collect many relevance judgments for a few topics -- i.e., {\em Narrow and Deep} (NaD) judging -- or a few relevance judgments for many topics -- i.e., {\em Wide and Shallow} (WaS) judging, for a given evaluation budget. For example, in the TREC Million Query Track \cite{carterette2009if}, IR systems were run on $\sim$10K queries sampled from two large query logs, and shallow judging was performed for a subset of topics for which a human assessor could ascribe some intent to the query such that a topic description could be back-fit and relevance determinations could be made.
Intuitively, since people search for a wide variety of topics expressed using a wide variety of queries, it makes sense to evaluate systems across a similarly wide variety of search topics and queries. Empirically, large variance in search accuracy is often observed for the same system across different topics~\cite{banks99}, motivating use of many diverse topics for evaluation in order to achieve stable evaluation of systems. 
Prior studies have reported a fairly consistent finding that WaS judging tends to provide more stable evaluation for the same human effort vs.\ NaD judging~\cite{sanderson2005information,carterette2007hypothesis,bodoff2007test}. While this finding does not hold in all cases, exceptions have been fairly limited. For example, \citeauthor{carterette2008evaluation}~\citeyear{carterette2008evaluation} achieve the same reliability using 250 topics with 20 judgments  per topic (5000 judgments in total)  as  600 topics with 10 judgments per topic (6000 judgments in total).
A key observation we make in this work is noting that all prior studies comparing NaD vs.\ WaS judging assume that search topics are selected randomly.

Another direction of research has sought to carefully choose which search topics are included in a test collection (i.e.,  {\em intelligent topic selection}) so as to minimize the number of search topics needed for a stable evaluation. Since human relevance judgments must be collected for any topic included, using fewer topics directly reduces judging costs. NIST TREC test collections have traditionally used 50 search topics (manually selected from a larger initial set of candidates), following a simple, effective, but costly topic creation process which includes collecting initial judgments for each candidate topic and manual selection of final topics to keep~\cite{voorhees2001philosophy}.  \citeauthor{BuckleyVoorhees2000}~\citeyear{BuckleyVoorhees2000} report that at least 25 topics are needed for stable evaluation, with 50 being better, while \citeauthor{Zobel1998}~\citeyear{Zobel1998} showed that one set of 25 topics predicted relative performance of systems fairly well on a different set of 25 topics. \citeauthor{guiver2009few}~\citeyear{guiver2009few} conducted a systematic study showing that evaluating IR systems using the ``right'' subset of topics yields very similar results vs.\ evaluating systems over all topics. However, they did not propose a method to find such an effective topic subset in practice. Most recently,  \citeauthor{hosseini2012uncertainty}~\citeyear{hosseini2012uncertainty} proposed an iterative algorithm to find effective topic subsets, showing encouraging results. A key observation we make is that prior work on intelligent topic selection has not evaluated against shallow judging baselines, which tend to be the preferred strategy today for reducing IR evaluation cost. We argue that one must ask whether it is actually useful to select topics, or should one simply perform WaS judging over many topics? 

{\bf Our Work.} In this article, we propose a new {\em intelligent topic selection} method which reduces the number of search topics (and thereby costly human relevance judgments) needed for reliable IR evaluation. To rigorously assess our over-arching question of whether topic selection is actually useful in comparison to WaS judging approaches, we integrate previously disparate lines of research on intelligent topic selection and NaD vs.\ WaS judging. Specifically, we investigate a comprehensive set of relevant factors never previously considered together: 1) method of topic selection; 2) the effect of topic familiarity on human judging speed; and 3) how different topic generation processes (requiring varying human effort) impact (i) budget utilization and (ii) the resultant quality of judgments.
We note that prior work on NaD vs.\ WaS judging has not considered cost ramifications of how judging depth impacts judging speed (i.e., assessors becoming faster at judging a particular topic as they become more familiar with it). Similarly, prior work on NaD vs.\ WaS judging has not considered topic construction time; WaS judging of many topics appears may be far less desirable if we account for traditional NIST TREC topic construction time \cite{ellenvoorheesemail}. As such, our findings also further inform the broader debate on NaD vs.\ WaS judging assuming random topic selection.

We begin with our first research question~\textbf{RQ-1}: {\em How can we select search topics that maximize evaluation validity given document rankings of multiple IR systems for each topic?}
We propose a novel application of \emph{learning-to-rank} (L2R) to topic selection. In particular, topics are selected iteratively via a greedy method  which optimizes accurate ranking of systems (Section~\ref{our_greedy_approach}). We adopt MART~\cite{friedman2001greedy} as our L2R model, though our approach is largely agnostic and other L2R models might be used instead. We define and extract 63 features for this topic selection task which represent the interaction between topics and ranking of systems (Section~\ref{features_section}). 
To train our model, we propose a method to automatically generate useful training data from existing test collections (Section~\ref{generate_training_data}). By relying only on pre-existing test collections for model training, we can construct a new test collection without any prior relevance judgments for it, rendering our approach more generalizable and useful. We evaluate our approach on NIST TREC Robust 2003~\cite{voorhees2003overview} and Robust 2004~\cite{voorhees2004overview} test collections, 
%
%
comparing our approach to recent prior work~\cite{hosseini2012uncertainty} and \emph{random} topic selection (Section~\ref{experiments}). 
Results show consistent improvement over baselines, with greater relative improvement as fewer topics are used.


In addition to showing improvement of our topic selection method over prior work, as noted above, we believe it is essential to assess intelligent topic selection in regard to the real over-arching question: what is the best way to achieve cost-effective IR evaluation? Is intelligent topic selection actually useful, or should we simply do WaS judging over many topics? To investigate this, we conduct a comprehensive analysis involving a set of  focused research questions not considered by the prior work, all utilizing our intelligent topic selection method:

\begin{itemize}

\item \textbf{RQ-2: When topics are selected intelligently, and other factors held constant, is WaS judging (still) a better way to construct test collections than NaD judging?} 
When intelligent topic selection is used, we find that NaD judging often achieves greater evaluation reliability than WaS judging for the same budget when topics are selected intelligently, contrasting with popular wisdom today favoring WaS judging.



\item {\bf RQ-3 (Judging Speed and Topic Familiarity): Assuming WaS judging leads to slower judging speed than NaD judging, how does this impact our assessment of intelligent topic selection?} Past comparisons between NaD vs.\ WaS judging have typically assumed constant judging speed~\cite{carterette2007hypothesis,sanderson2005information}. However, data reported by \citeauthor{carterette2009if}~\cite{carterette2009if} suggests that assessors may judge documents faster as they judge more documents for the same topic (likely due to increased topic familiarity). 
Because we can collect more judgments in the same amount of time with NaD vs.\ WaS judging, we show that NaD judging achieves greater relative evaluation reliability than shown in prior studies, which did not consider the speed benefit of deep judging.

\item {\bf RQ-4 (Topic Development Time): How does topic development time in the context of NaD vs.\ WaS judging impact our assessment of intelligent topic selection?} Prior NaD vs.\ WaS studies have typically ignored non-judging costs involved in test collection construction. While \citeauthor{carterette2008evaluation}~\citeyear{carterette2008evaluation} consider topic development time, the 5-minute time they assumed is roughly two orders of magnitude less than the 4 hours NIST has traditionally taken to construct each topic~\cite{ellenvoorheesemail}. 
We find that WaS judging is preferable than NaD judging for short topic development times (specifically $\leq$5 minutes in our experiments).  However, as  the topic development cost further increases, NaD judging becomes increasingly preferable. 


\item \textbf{ RQ-5 (Judging Error): Assuming short topic development times reduce judging consistency, how does this impact our assessment of intelligent topic selection in the context of NaD vs.\ WaS judging?} Several studies have reported calibration effects impacting the decisions and consistency of relevance assessors ~\cite{sanderson2010relatively,scholer2013effect}. 
While NIST has traditionally included an initial ``burn-in'' judging period as part of topic generation and formulation~\cite{ellenvoorheesemail}, we posit that drastically reducing topic development time (e.g., from 4 hours~\cite{ellenvoorheesemail} to 2 minutes \cite{carterette2009if}) could negatively impact topic quality, leading to less well-defined topics and/or calibrated judges, and thereby less reliable judgments. As suggestive evidence, \citeauthor{McDonnell2016}~\citeyear{McDonnell2016} report high judging agreement in reproducing a ``standard'' NIST track, but high and inexplicable judging disagreement on TREC's Million Query track~\cite{carterette2009if}, which lacked any burn-in period for judges and had far shorter topic generation times. To investigate this, we simulate increased judging error as a function of lower topic generation times.
We find that it is better to invest a portion of our evaluation budget to increase quality of topics, instead of collecting more judgments for low-quality topics. This also makes NaD judging preferable in many cases, due to increased topic development cost.
\end{itemize}

{\bf Contributions.} Our five research questions address the over-arching goal and challenge of minimizing IR evaluation cost while ensuring validity of evaluation. Firstly, we propose an intelligent topic selection algorithm, as a novel application of learning-to-rank, and show its effectiveness vs.\ prior work. Secondly, we go beyond prior work on topic selection to investigate whether it is actually useful, or if one should simply do WaS judging over many topics rather than topic selection? Our comprehensive analysis over several factors not considered in prior studies  shows that intelligent topic selection is indeed useful, and contrasts current wisdom favoring WaS judging.


The remainder of this article is organized as follow. Section~\ref{related_work} reviews the related work on topic selection and topic set design. Section~\ref{problem_definition} formally defines the topic selection problem. In Section~\ref{proposed_approach}, we describe our proposed L2R-based approach in detail. Section~\ref{experiments} presents our experimental evaluation. Finally, Section~\ref{conclusion} summarizes the contributions of our work and suggests potential future directions.

\section{Related Work} 
\label{related_work}


Constructing test collections is expensive in human effort required. Therefore, researchers have proposed a variety of methods to reduce the cost of creating test collections. Proposed methods include: developing new evaluation measures and statistical methods for  the case of incomplete judgments~\cite{aslam2006statistical,buckley2004retrieval,sakai2007alternatives, yilmaz2006estimating, yilmaz2008estimating}, finding the best sample of documents to be judged for each topic~\cite{ cormack1998efficient,carterette2006minimal,jones1975report, moffat2007strategic, pavlu2007practical}, 
inferring relevance judgments~\cite{aslam2007inferring}, 
topic selection~\cite{hosseini2012uncertainty, hosseini2011selecting, mizzaro2007hits, guiver2009few}, evaluation with no human judgments~\cite{nuray2006automatic, soboroff2001ranking}, crowdsourcing~\cite{alonso2009can, grady2010crowdsourcing}, and others. The reader is referred to \cite{moghadasi2013low} and \cite{sanderson2010test} for a more detailed review of prior work on methods for low-cost IR evaluation. 



\subsection{Topic Selection} 
\label{sec:topic_selection}

To the best of our knowledge, \citeauthor{mizzaro2007hits}~\citeyear{mizzaro2007hits}'s study was the first seeking to select the best subset of topics for evaluation. They first built a system-topic graph representing the relationship between topics and IR systems,  then ran the HITS algorithm on it. They hypothesized that  topics with higher ‘hubness’ scores would  better distinguish between systems. However,  \citeauthor{robertson2011contributions}~\citeyear{robertson2011contributions} experimentally showed that their hypothesis was not true.

\citeauthor{guiver2009few}~\citeyear{guiver2009few} experimentally showed that if we choose the right subset of topics, we can achieve a ranking of systems that is very similar to the ranking when we employ all topics. However they did not provide a solution to find the right subset of topics. This study has motivated other researchers to investigate this problem. \citeauthor{berto2013using}~\citeyear{berto2013using} stressed  generality and showed that a carefully selected good subset of topics to evaluate  a set of systems  can be also adequate to evaluate a different set of systems. \citeauthor{hauff2010case}~\citeyear{hauff2010case} reported that using the easiest topics based on Jensen-Shannon Divergence  approach did not work well to reduce the number of topics. \citeauthor{hosseini2011selecting}~\citeyear{hosseini2011selecting} focused on selecting the subset of topics to extend an existing collection in order to increase its re-usability. \citeauthor{culpepper2014trec}.~\citeyear{culpepper2014trec} investigated how
the capability of topics to predict overall system effectiveness has changed over the years in TREC test collections.
\citeauthor{kazai2014dissimilarity}~\citeyear{kazai2014dissimilarity} reduced the cost using dissimilarity based query selection for preference based IR evaluation.


The closest study to our own is \cite{hosseini2012uncertainty}, which employs an adaptive algorithm for topic selection. It selects the first topic randomly. Once a topic is selected, the relevance judgments are acquired and used to assist with the selection of subsequent topics. 
Specifically, in the following iterations, the topic that is predicted to maximize the current Pearson correlation is selected. In order to do that, they predict relevance probabilities of \emph{qrels} for the remaining topics using a Support Vector Machine (SVM) model trained on the judgments from the topics selected thus far. Training data is extended at each iteration by adding the relevance judgments from each topic as it is selected in order to better select the next topic. 

Further studies investigated topic selection for other purposes, such as creating low-cost datasets for training learning-to-rank algorithms~\cite{mehrotra2015representative}, system rank estimation~\cite{hauff2009relying}, and selecting training data to improve supervised data fusion algorithms~\cite{lin2011query}. These studies do not consider topic selection for low-cost evaluation of IR systems. 

\subsection{How Many Topics Are Needed?} 
\label{sec:topic_set_size_design}

Past work has investigated the ideal size of test collections and how many topics are needed for a reliable evaluation. While traditional TREC test collections employ 50 topics, a number of researchers claimed that 50 topics are not sufficient for a reliable evaluation~\cite{jones1975report,voorhees2009topic, urbano2013measurement,sakai2015topic}. Many researchers reported that wide and shallow 
 judging is preferable than narrow and deep judging~\cite{sanderson2005information,carterette2007hypothesis,bodoff2007test}. 
 \citeauthor{carterette2008evaluation}~\citeyear{carterette2008evaluation} experimentally compared deep vs. shallow judging in terms of budget utilization. They found that 20 judgments with 250 topics was the most cost-effective in their experiments. \citeauthor{urbano2013measurement}~\citeyear{urbano2013measurement} measured the reliability of TREC test collections with regard to generalization and concluded that the number of topics needed for a reliable evaluation varies across different tasks. \citeauthor{Urbano2016}~\citeyear{Urbano2016} analyzed different test collection reliability measures with a special focus on the number of topics.

In order to calculate the number of topics required, \citeauthor{webber2008statistical}~\citeyear{webber2008statistical} proposed adding  topics iteratively until desired statistical power is reached. Sakai proposed  methods based on two-way ANOVA~\citeyear{sakai2014topic}, confidence interval~\citeyear{sakai2014CI}, and \emph{t} test and one-way ANOVA~\citeyear{sakai2014designing}. In his follow-up studies, Sakai investigated the effect of score standardization~\citeyear{sakai2016simple} in topic set design \mbox{\citeyear{Sakai2016}}   
and provided guidelines for test collection design for a given fixed budget \citeyear{sakai2015topic}. \citeauthor{Sakai2015}~\citeyear{Sakai2015} applied the method of \citeauthor{sakai2015topic}~\citeyear{sakai2015topic} to decide the number of topics for evaluation measures of a Short Text Conversation task\footnote{http://ntcir12.noahlab.com.hk/stc.htm}. \citeauthor{sakai2016onestimating} \citeyear{sakai2016onestimating} explored how many topics and IR systems are needed for a reliable topic set size estimation.  
While these studies focused on calculating the number of topics required, our work focuses on how to select the best topic set for a given size in order to maximize the reliability of evaluation. 
We also investigate further considerations impacting the debate over shallow vs.\  deep judging: familiarization of users to topics, and the effect of topic development costs on the budget utilization and the  quality of judgments for each topic.

\citeauthor{ASI:ASI23304}~\citeyear{ASI:ASI23304} investigated the problem of evaluating commercial search engines by sampling queries based on their distribution in query logs. In contrast, our work does not rely on any prior knowledge about the popularity of topics in performing topic selection.



\subsection{Topic Familiarity vs.\ Judging Speed} 
\label{sec:topic_familiarity}

\citeauthor{carterette2009if}~\citeyear{carterette2009if} reported that as the number of judgments per topic increases (when collecting 8, 16, 32, 64 or 128 judgments per topic), the median time to judge each document decreases respectively: 15, 13, 15, 11 and 9 seconds. This suggests that assessors become more familiar with a topic as they judge more documents for it, and this greater familiarity yields greater judging speed. However, prior work comparing deep vs.\ shallow judging did not consider this, instead assuming that judging speed is constant regardless of judging depth. Consequently, our experiments in Section~\ref{selecting_with_fixed_budget} revisit this question, considering how faster judging with greater judging depth per topic may impact the tradeoff between deep vs.\ shallow judging in maximizing evaluation reliability for a given budget in human assessor time.

\subsection{Topic Development Cost vs.\ Judging Consistency} 
\label{sec:topic_generation_cost}

Past work has utilized a variety of different processes to develop search topics when constructing test collections. These different processes explicitly or implicitly enact potentially important trade-offs between human effort (i.e.\ cost) vs.\ quality of the resultant topics developed by each process. For example, NIST has employed a relatively costly process in order to ensure creation of very high quality topics~\cite{ellenvoorheesemail}:

\begin{quotation}
\noindent For the traditional ad hoc tasks, assessors generally came to NIST with some rough ideas for topics having been told the target document collection.  For each idea, they would create a query and judge about 100 documents (unless at least 20 of the first 25 were relevant, in which case they would stop at 25 and discard the idea).  From the set of candidate topics across all assessors, NIST would select the final test set of 50 based on load-balancing across assessors, number of relevant found,  eliminating duplication of subject matter or topic types, etc.  The judging was an intrinsic part of the topic development routine because we needed to know that the topic had sufficiently many (but not too many) relevant in the target document set.  (These judgments made during the topic development phase were then discarded.  Qrels were created based only on the judgments made during the official judgment phase on pooled participant results.)  We used a heuristic that expected one out of three original ideas would eventually make it as a test set topic.  Creating a set of 50 topics for a newswire ad hoc collection was budgeted at about 175-225 assessor hours, which works out to about 4 hours per final topic.
\end{quotation}

In contrast, the TREC Million Query (MQ) Track used a rather different procedure to develop topics. In the 2007 MQ Track~\cite{allan2007million}, 10000 queries were sampled from a large search engine query log.
The assessment system showed 10 randomly selected queries to each assessor, who then selected one and converted it into a standard TREC topic by back-fitting a topic description and narrative to the selected query. \citeauthor{carterette2008evaluation}~\citeyear{carterette2008evaluation} reported that the median time of developing a topic was roughly 5 minutes. In the 2008 MQ Track~\cite{Allan08millionquery}, assessors could refresh list of candidate 10 queries if they did not want to judge any of the candidates listed. \citeauthor{carterette2009if}~\citeyear{carterette2009if} reported that \emph{median} time for viewing a list of queries was 22 seconds and back-fitting a topic description was 76 seconds. On average, each assessor viewed 2.4 lists to develop each topic. Therefore, the cost of developing a topic was roughly $2.4*22 + 76 \approx 129$ seconds, or 2.1 minutes.

The examples above show a vast range of topic creation times: from 4 hours to 2 minutes per topic. Therefore, in Section~\ref{selecting_with_fixed_budget}, we investigate deep vs.\  shallow judging when cost of developing topics is also considered.

In addition to considering topic construction time, we might also consider whether aggressive reduction in topic creation time might also have other unintended, negative impacts on topic quality. For example, \citeauthor{scholer2013effect}~\citeyear{scholer2013effect} reported calibration effects change judging decisions as assessors familiarize themselves with a topic. Presumably NIST's 4 hour topic creation process provides judges ample time to familiarize themselves with a topic, and as noted above, judgments made during the topic development phase are then discarded. In contrast, it seems MQ track assessors began judging almost immediately after selecting a query for which to back-fit a topic, and with no initial topic formation period for establishing the topic and discarding initial topics made during this time. Further empirical evidence suggesting quality concerns with MQ track judgments was also recently reported by \citeauthor{McDonnell2016}~\citeyear{McDonnell2016}, who described a detailed judging process they employed to reproduce NIST judgments. While the authors reported high agreement between their own judging and crowd judging vs.\ NIST on the 2009 Web Track, for NIST judgments from the 2009 MQ track, the authors and crowd judges were both consistent while disagreeing often with NIST judges. The authors also reported that even after detailed analysis of the cases of disagreement, they could not find a rationale for the observed MQ track judgments. Taken in sum, these findings suggest that aggressively reducing topic creation time may negatively impact the quality of judgments collected for that topic. For example, while an assessor is still formulating and clarify a topic for himself/herself, any judgments made at this early stage of topic evolution may not be self-consistent with judgments made once the topic is further crystallized. Consequently, in Section~\ref{selecting_with_fixed_budget} we revisit the question of deep judging of few topics vs.\ shallow judging of many topics, assuming that low topic creation times may  also mean less consistent judging.





\section{Problem Definition}\label{problem_definition}

In this section, we define the topic selection problem.
We assume that we have a TREC-like setup: a document collection has already been acquired, a large pool of topics and ranked  lists of IR systems for each topic are also available. Our goal is to select a certain number of topics from the topic pool such that evaluation with those selected topics yields the most similar ranking of the IR systems to the ``\emph{ground-truth}''. 
 We assume that the ground-truth ranking of the IR systems is the one when we use all topics in the pool for evaluation. 

We can formulate this problem as follows.
Let $T=\{t_1,t_2,...,t_N\}$ denote the pool of $N$ topics, $S=\{s_1,s_2,...,s_K\}$ denote the set of $K$ IR systems to be evaluated, and $R_{<S,T,e>}$ denote the ranking of systems in $S$ when they are evaluated based on evaluation measure $e$ over the topic set $T$ (notation used in equations and algorithms is shown in \textbf{Table~\ref{notation_table}}). 
  We aim to select a subset $P \subset T$ of $M$ topics that \emph{maximizes} the correlation (as a measure of similarity between two ranked lists) between the ranking of systems over $P$ (i.e., considering only $M$ topics and their corresponding relevance judgments) and the ground-truth ranking of systems (over $T$). Mathematical definition of our goal is as follows:
\begin{equation}\label{equation_maximize}
\underset{P\subset T,\ |P|=M}{max} corr(R_{<S,P,e>}, R_{<S,T,e>} )
\end{equation}

\noindent
where \emph{corr} is a ranking similarity function, {\it e.g.,} Kendall-$\tau$~\cite{kendall1938new}.

\begin{table*}[t]
\centering
\caption{Notation used in equations and algorithms}
\vspace{5pt}
\label{notation_table}
    \begin{tabular}{| p{1.5cm} | p{10cm}   | } \hline
   		\textbf{Symbol} & \textbf{Name}  \\ \hline
        $T$ & Topic pool   \\ \hline
        $S$ & IR systems participated to the pool of the corresponding test collection   \\ \hline
        $R_{<S,T,e>}$ & Ranking of systems in $S$ when they are evaluated based on evaluation measure $e$ over the topic set $T$   \\ \hline
        $N$ & Size of topic pool   \\ \hline
        $M$ & Number of topics to be selected   \\ \hline
        $D_{t_{c}}$ & The document pool for topic $t_c$   \\ \hline
      	$L_{s_j(t_c)}$ & The ranked list resulting from system $s_j$ for the topic $t_c$   \\ \hline
    
 	\end{tabular}  
\end{table*}

\section{Proposed Approach} \label{proposed_approach}
The problem we are tackling is challenging since we do not know the actual performance of systems (i.e.\  their performance when all topics are employed for evaluation) and we would like to find a subset of topics that achieves similar ranking to the \textit{unknown} ground-truth. 

To demonstrate the complexity of the problem, let us assume that we obtain the judgments for all topic-document pairs (i.e.,\  we know the ground-truth ranking). In this case, we have to check $N \choose M$ possibilities of subsets in order to find the optimal one (i.e., the one that produces a ranking that has the maximum correlation with the ground-truth ranking). For example, if $N=100$ and $M=50$, we need to check around $10^{29}$ subsets of topics. Since this is computationally intractable, we need an approximation algorithm to solve this problem. Therefore, we first describe a greedy oracle approach to select the best subset of topics when we already have the judgments for all query-document pairs (Section~\ref{greedy_approach}). Subsequently, we discuss how we can employ this greedy approach when we do not already have  the relevance judgments (Section~\ref{prediction_based_approach}). Finally, we introduce our L2R-based topic selection approach (Section~\ref{our_greedy_approach}). 
\subsection{Greedy Approach}\label{greedy_approach}
We first explore a greedy oracle approach that selects topics in an iterative way when relevance judgments are already obtained. Instead of examining all possibilities, at each iteration, we select the '\emph{best}' topic (among the currently non-selected ones) that, when added to the currently-selected subset of topics, will produce the ranking that has the maximum correlation with the ground-truth ranking of systems. 
 
 \textbf{Algorithm~\ref{greedy_algorithm}} illustrates this oracle greedy approach. First, we initialize set of selected topics (P) and set of candidate topics to be selected ($\bar{P}$) (Line 1). 
 For each candidate topic $t$ in $\bar{P}$, we rank the systems over the selected topics $P$ in addition to $t$ ($R_{<S,P\cup \{t\},e>}$), and calculate the Kendall's $\tau$ achieved with this ranking (Lines 3-4). We then pick the topic achieving the highest Kendall-$\tau$ score among other candidates (Line 5) and update $P$ and $\bar{P}$ accordingly (Lines 6-7). We repeat this process until we reach the targeted subset size $M$ (Lines 2-7).   

\begin{algorithm}[H]
	\caption{A Greedy Oracle  Approach for Topic Selection}
	\label{greedy_algorithm}
	\begin{algorithmic}[1]
		\State $P \leftarrow \emptyset$; $\bar{P} \leftarrow T$ \Comment{ selected set $P$ is initially empty}
        \While {$|P| < M$ }
        	\ForEach {topic $t \in \bar{P}$}
                \State $\tau_t \leftarrow corr(R_{<S,P\cup \{t\},e>},\ R_{<S,T,e>})$
            \EndFor
            \State topic $t^* \leftarrow \underset{t \in \bar{P}}{max} (\tau_t)$ \Comment{choose $t^*$ yielding the best correlation}
            \State $P \leftarrow P\cup \{t^*\}$  \Comment{add $t^*$ to the selected set $P$}
            \State $\bar{P} \leftarrow \bar{P}-\{t^*\}$
		\EndWhile
		\vspace{0.13cm}
	\end{algorithmic}
\end{algorithm}


While this approach has $O(M\times N)$ complexity (which is clearly much more efficient compared to selecting the optimal subset), it is also impractical due to leveraging the real judgments (which we typically do not have in advance) in order to calculate the ground-truth ranking and thereby Kendall-$\tau$ scores.

\subsection{Performance Prediction Approach}\label{prediction_based_approach}
One possible way to avoid the need for the actual relevance judgments is to predict the performance of IR systems using automatic evaluation methods~\cite{soboroff2001ranking, nuray2006automatic} and then rank the systems based on their predicted performance. 
For example, \citeauthor{hosseini2012uncertainty}~\citeyear{hosseini2012uncertainty} predict relevance probability of document-topic pairs by employing an SVM classifier and select topics in a greedy way similar to  Algorithm~\ref{greedy_algorithm}. We use their selection approach as a baseline in our experiments (Section~\ref{experiments}).


\subsection{Proposed Learning-to-Rank Approach}\label{our_greedy_approach}
In this work, we formulate the topic selection problem as a learning-to-rank (L2R) problem. In a typical L2R problem, we are given a query $q$ and a set of documents $D$, and a model is learned to rank those documents in terms of relevance with respect to $q$. The model is trained using a set of queries and their corresponding labeled documents. 
 In our context, we are given a set of currently-selected topic set $P$ (analogous to the query $q$) and the set of candidate topics $\bar{P}$ to be selected from (analogous to the documents $D$), and we aim to train a model to rank the topics in $\bar{P}$ based on the expected effectiveness of adding each to $P$. 
The training samples used to train the model are tuples of the form $(P, t, corr(R_{<S, P\cup\{t\}, e>}, R_{<S, T, e>}))$, where the measured correlation is used to label  topic $t$ with respect to $P$. Notice that the correlation is computed using the true relevance judgments in the training data. This enables us to use the wealth of existing test collections to acquire data for training our model, as explained in Section~\ref{generate_training_data}.

We apply this L2R problem formulation to the topic selection problem using our greedy approach. 
We use the trained L2R model to rank the candidate topics and then select the first-ranked one. The algorithm is shown in \textbf{Algorithm~\ref{algorithm_our_method}}. At each iteration, a feature vector $v_t$ is computed for each candidate topic $t$ in $\bar{P}$  using a feature extraction function $f$ (Lines 3-4), detailed in Section~\ref{features_section}. The candidate topics are then ranked using our learned model (Line 5) and the topic in the first rank is picked (Line 6). Finally, the topic sets $P$ and $\bar{P}$ are updated (Lines 7-8) and a new iteration is started, if necessary.  


\begin{algorithm}[H]
	\caption{L2R-based Topic Selection}
	\label{algorithm_our_method}
	\begin{algorithmic}[1]
    		\State $P \leftarrow \emptyset$; $\bar{P} \leftarrow T$ \Comment{ selected set $P$ is initially empty}
        \While {$|P| < M$ }
        	\ForEach {topic $t \in \bar{P}$}
            	\State $v_t \leftarrow f(t,P,\bar{P})$ \Comment{compute feature vector for $t$}
            \EndFor
            \State $l \leftarrow L2R(\{(t, v_t)\}\ |\ \forall t\in \bar{P})$ \Comment{apply L2R on feature vectors}
            \State topic $t^* \leftarrow first(l)$ \Comment{choose $t^*$ in the first rank}
            \State $P \leftarrow P\cup \{t^*\}$ \Comment{add $t^*$ to the selected set $P$}
            \State $\bar{P} \leftarrow \bar{P}-\{t^*\}$ \Comment{Remove $t^*$ from the candidate set $\bar{P}$}
		\EndWhile
		\vspace{0.13cm}
	\end{algorithmic}
\end{algorithm}

\subsubsection{Features}\label{features_section}
In this section, we describe the features we extract in our L2R approach for each candidate topic. 
\citeauthor{hosseini2012uncertainty}~\citeyear{hosseini2012uncertainty} mathematically show that, in the greedy approach, the topic selected  at each iteration should be different from the already-selected ones (i.e.,\ topics in P) while being  representative of the non-selected ones (i.e.,\ topics in  $\bar{P}$). Therefore, the extracted set of features should cover the candidate topic as well as the two sets $P$ and $\bar{P}$. Features should therefore capture the interaction between the topics and the IR systems in addition to the diversity between the IR systems in terms of their retrieval results. 

We define two types of feature sets. \emph{Topic-based} features are extracted from an individual topic while \emph{set-based} features are extracted from a set of topics by aggregating the topic-based features extracted from each of those topics. 

The topic-based features include 7 features that are extracted for a given candidate topic $t_c$ and are listed in \textbf{Table~\ref{features_table}}. 
For a given set of topics (e.g., currently-selected topics $P$), we extract the set-based features by computing both average and standard deviation of each of the 7 topic-based features extracted from all topics in the set. This gives us 14 set-based features that can be extracted for a set of topics. We compute these 14 features for each of the following sets of topics:
\begin{itemize}
\item currently-selected topics ($P$)
\item not-yet-selected topics ($\bar{P}$)
\item selected topics with the candidate topic ($P\cup \{t_c\}$)
\item not-selected topics excluding the candidate topic ($\bar{P}-\{t_c\}$)
\end{itemize}

\begin{table*}[t]
\centering
\caption{Topic-based Features}
\vspace{5pt}
\label{features_table}
    \begin{tabular}{| p{1.5cm} | p{10cm}   | } \hline
   		\textbf{Feature} & \textbf{Description}  \\ \hline
        $f_{\bar{w}}$ & Average sampling weight of documents   \\ \hline
        $f_{\sigma_w}$ & Standard deviation of weight of documents   \\ \hline
        $f_{\bar{\tau}}$ & Average $\tau$ score for ranked lists pairs   \\ \hline
        $f_{\sigma_\tau}$ & Standard deviation of $\tau$ scores for ranked lists pairs   \\ \hline
        $f_\$$ & Judgment cost of the topic   \\ \hline
        $f_{\sigma_\$}$ & Standard deviation of judgment costs of system pairs   \\ \hline
        $f_{\sigma_{QPP}}$ & Standard deviation of estimated performance of systems   \\ \hline
 	\end{tabular}  
\end{table*}

In total, we have 63 features for each data record representing a candidate topic: $14 \times 4 = 56$ features for the above groups + 7 topic-based features. We now describe the seven topic-based features that are at the core of the  feature set. 

\begin{itemize}
\item 
\textbf{Average sampling weight of documents ($f_{\bar{w}}$)}: In the statAP sampling method~\cite{pavlu2007practical}, a weight is computed for each document based on where it appears in the ranked lists of all IR systems. Simply, the documents at higher ranks get higher weights. The weights are then used in a non-uniform sampling strategy to sample more documents relevant to the corresponding topic. We compute the average sampling weight of all documents that appear in the pool of the candidate topic $t_c$ as follows: 
\begin{equation}
\label{f_ave_weight}
f_{\bar{w}}(t_c) = \frac{1}{|D_{t_c}|} \sum_{d \in D_{t_c}} w(d,S)
\end{equation}
\noindent
where $D_{t_c}$ is the document pool for topic $t_c$ and $w(d, S)$ is the weight of document $d$ over the IR systems $S$. High $f_{\bar{w}}$ values mean that the systems have common documents at higher ranks for the corresponding topic, whereas lower $f_{\bar{w}}$ values indicate that
the systems return significantly different ranked lists or have only the documents at lower ranks  in common.

\item 
\textbf{Standard deviation of weight of documents ($f_{\sigma_w}$)}: Similar to $f_{\bar{w}}$, we also compute the standard deviation of the sampling weights of documents for the candidate topic as follows: 
\begin{equation}
\label{f_std_weight}
f_{\sigma_w}(t_c) = \sigma\{w(d,S)\ |\ \forall d \in D_{t_c}\} 
\end{equation}

\item 
\textbf{Average $\tau$ score for ranked lists pairs ($f_{\bar{\tau}}$)}: 
This feature computes Kendall's $\tau$ correlation between ranked lists of each pair of the IR systems and then takes the average (as shown in Equation~\ref{f_ave_tau}) in order to capture the diversity of the results of the IR systems. The depth of the ranked lists is set to 100. In order to calculate the Kendall's $\tau$ score, the documents that appear in one list but not in the other are concatenated to the other list so that both ranked lists contain the same documents. If there are multiple documents to be concatenated, the order of the documents in the ranked list is preserved during concatenation. For instance, if system \textcolor{blue}{B} returns documents \{\textcolor{blue}{a},\textcolor{blue}{b},\textcolor{blue}{c},\textcolor{blue}{d}\} and system \textcolor{red}{R} returns \{\textcolor{red}{e},\textcolor{red}{a},\textcolor{red}{f},\textcolor{red}{c}\} for a topic, then the concatenated ranked lists of \textcolor{blue}{B} and \textcolor{red}{R} are \{\textcolor{blue}{a},\textcolor{blue}{b},\textcolor{blue}{c},\textcolor{blue}{d},\textcolor{red}{e},\textcolor{red}{f}\} and \{\textcolor{red}{e},\textcolor{red}{a},\textcolor{red}{f},\textcolor{red}{c},\textcolor{blue}{b},\textcolor{blue}{d}\}, respectively. 
\begin{equation} \label{f_ave_tau}
f_{\bar{\tau}}(t_c) =\frac{1}{2|S|-1} \sum_{i=1}^{|S|-1} \sum_{j=i+1}^{|S|} corr(L_{s_i(t_c)}, L_{s_j(t_c)}) 
\end{equation}
\noindent
where $L_{s_j(t_c)}$ represents the ranked list resulting from system $s_j$ for the topic $t_c$.

\item 
\textbf{Standard deviation of $\tau$ scores for ranked lists pairs ($f_{\sigma_\tau}$)}: This feature computes the standard deviation of the $\tau$ scores of the pairs of ranked lists as follows:

\begin{equation}
\label{f_std_tau}
f_{\sigma_\tau}(t_c) =\sigma \{corr(L_{s_i(t_c)}, L_{s_j(t_c)})\ |\ \forall i,j \leq |S|, i \neq j\}
\end{equation}

\item 
\textbf{Judgment cost of the topic ($f_\$$)}: This feature estimates the cost of judging the candidate  topic as the number of documents in the pool at a certain depth. If IR systems return many different documents, then the judging cost increases; otherwise, it decreases due to having many documents in common. We set pool depth to 100 and normalize costs by dividing by the maximum possible cost (i.e., $100 \times |S|$). 

\begin{equation}
\label{f_cost}
f_\$(t_c) = \frac{D_{t_c}}{|S| \times 100} 
\end{equation}

\item 
\textbf{Standard deviation of judgment costs of system pairs ($f_{\sigma_\$}$)}: The judgment cost depends on systems participating in the pool. We construct the pool separately for each pair of systems and compute the standard deviation of the judgment cost across pools as follows: 

\begin{equation}
\label{f_std_cost}
f_{\sigma_\$}(t_c) =\sigma \{|L_{s_i(t_c)} \cup L_{s_j(t_c)}|\ |\ \forall i,j \leq |S|, i \neq j\}
\end{equation}

\item 
\textbf{Standard deviation of estimated performance of systems ($f_{\sigma_{QPP}}$)}: We finally compute standard deviation of the estimated performances of the IR systems for the topic $t_c$ using a  query performance predictor (QPP)~\cite{cummins2011improved}. 
The  QPP is typically used to estimate the performance of a single system and  is affected by the range of  retrieval scores of retrieved documents. Therefore, we normalize the document scores using min-max normalization before computing the predictor.    
\begin{equation}
\label{f_std_performance}
f_{\sigma_{QPP}}(t_c) = \sigma\{QPP(s_i,t_c)\ |\ \forall i \leq |S|\}
\end{equation}
\noindent
where $QPP(s_i,t_c)$ is the performance predictor applied 
on system $s_i$ given topic $t_c$.
\end{itemize}

\subsubsection{Generating Training Data} 
\label{generate_training_data}

Our proposed L2R approach ranks  topics based on their effectiveness when added to some currently-selected set of topics. This makes creating the training data for the model a challenging task. First, there are countless number of possible scenarios ({\it i.e.}, different combinations of topic sets) that we can encounter during the topic selection process. Second, the training data should specify which topic is more preferable for a given scenario. 

We developed a method to generate training data by leveraging existing test collections for which we have both relevance judgments and document rankings from several IR systems ({\it e.g.}, TREC test collections). We first simulate a scenario in which a subset of topics has already been selected.  We then rank the rest of the topics based on the correlation with the ground-truth ranking when each topic is added to the currently-selected subset of topics. We repeat this process multiple times and vary the number of already-selected topics in order to generate more diverse training data. The algorithm for generating training data from one test collection is given in \textbf{Algorithm~\ref{algorithm_generate_training_data}}.  The algorithm could also be applied to several test collections in order to generate larger training data.
\begin{algorithm}[H]
	\caption{Generating training data from one test collection of $T$ topics and $S$ IR systems}
	\label{algorithm_generate_training_data}
	\begin{algorithmic}[1]
    	\State $R^* \leftarrow R_{<S, T, e>}$ \Comment{Determine the ground-truth ranking of systems}
		\For {i = 0 to $|T|$ - 2}	\Comment{Vary the size of already-selected topic set}
        	\For {j = 1 to W} \Comment{Repeat the process $W$ times}
            	\State $P \leftarrow random (T,i)$ \Comment{Randomly select $i$ topics}
                \State $\bar{P} \leftarrow T - P$
                \ForEach {topic $t \in \bar{P}$}
                	\State $R_t  \leftarrow  R_{<S, P \cup t, e>}$ 
                    \State $\tau_t \leftarrow  corr(R_t, R^*)$
                 \EndFor 
                \State $\tau_{max} \leftarrow  \underset{t \in \bar{P}}{max} (\tau_t)$; $\tau_{min} \leftarrow  \underset{t \in \bar{P}}{min} (\tau_t)$
                \ForEach {topic $t \in \bar{P}$}
                	\State $v_t \leftarrow f(t,P,\bar{P})$ \Comment{Compute feature vector for $t$}  
                    \State $l_t \leftarrow \lfloor \frac{K * (\tau_t - \tau_{min})}{(\tau_{max} - \tau_{min})}\rfloor$   \Comment{Find which partition $\tau_t$  is in.}
                	\State Output  $<l_t,v_t >$
                \EndFor 
                
            \EndFor
        \EndFor
		\vspace{0.13cm}
	\end{algorithmic}
\end{algorithm}

The algorithm first determines the ground-truth ranking of IR systems using all topics in the test collection (Line 1). It then starts the process of generating the data records for each possible topic subset size for the targeted test collection (Line 2). 
For each subset size $i$, we repeat the following procedure $W$ times (Line 3); in each, we randomly select $i$ topics, assuming that these represent the currently-selected subset of topics $P$ (Line 4). For each topic $t$ of the non-selected topics $\bar{P}$, we rank the systems in case we add $t$ to $P$ and calculate the Kendall's $\tau$ score achieved in that case (Lines 6-9). This gives us how effective each of the candidate topics would be in the IR evaluation for this specific scenario ({\em i.e.}, when those $i$ topics are already selected). This also allows us to make a comparison between topics and rank them in terms of their effectiveness. In order to generate labels that can be used in L2R methods, we map each $\tau$ score to a value within a certain range. We first divide the range between maximum and minimum $\tau$ scores into $K$ equal bins and then assign each topic to its corresponding bin based on its effectiveness. For example, let $K=10$, $T_{max}= 0.9$, and $T_{min}=0.7$. The $\tau$ ranges for labeling will be $0=[0.7-0.72)$, $1=[0.72-0.74)$, ..., $9=[0.88-0.9]$. Topics are then labeled from 0 to ($K-1$) based on their assigned bin. For example, if we achieve  $\tau=0.73$ score for a particular topic, then the label for the corresponding data record will be 1. Finally, we compute the feature vector for each topic, assign the labels, and output the data records for the current scenario (Lines 10-13). We repeat this process $W$ times (Line 3) to capture more diverse scenarios for the given topic subset size. We can further increase the size of the generated training data by applying the algorithm on different test collections and merging the resulting data.

\section{Evaluation}
\label{experiments}

In this section, we evaluate our proposed L2R topic selection approach with respect to our research questions and baseline methods. Section~\ref{experimental_setup} details our experimental setup, including generation of training data and tuning of our L2R model parameters.  
We present results of our topic selection experiments (RQ-1) in Section~\ref{select_fix_number_of_topics}. We report ablation analysis of our features in Section~\ref{sec_feature_analysis} and discuss the evaluation of the parameters of our approach in Section~\ref{analysis_of_parameters}. In Section~\ref{selecting_with_fixed_budget}, we report the results of our experiments for intelligent topic selection  with a fixed budget (RQ-2) and considering different parameters in the debate of NaD vs.\ WaS judging: varying judging speed (RQ-3), topic generation time (RQ-4), and judging error (RQ-5). 


\subsection{Setup}\label{experimental_setup}
We adopt the MART~\cite{friedman2001greedy} implementation in the RankLib library\footnote{\url{https://sourceforge.net/p/lemur/wiki/RankLib/}} as our L2R model\footnote{We also focused on other L2R models but MART yielded the best results in our initial experiments when developing our method.}. To tune MART parameters, we partition our data into disjoint training, tuning, and testing sets. We assume that the ground-truth ranking of systems is given by MAP@100. 

\textbf{Test Collections}. We consider two primary criteria in selecting test collections to use in our experiments: (1) the collection should contain many topics, providing a fertile testbed for topic selection experimentation, and (2) the set of topics used in training, tuning, and testing should be disjoint to avoid over-fitting. 
To satisfy these criteria, we adopt the TREC-9~\cite{hawking2000overview} and TREC-2001~\cite{hawking2002overview} Web Track collections, as well as TREC-2003~\cite{voorhees2003overview} and TREC-2004~\cite{voorhees2004overview} Robust Track collections. Details of these test collections are presented in Table~\ref{datasets}. Note that all four collections target ad-hoc retrieval. We use TREC-9 and TREC-2001 test collections to generate our training data. 

\begin{table*}[htb]
\centering
\caption{Test Collections Used in Experiments}
\vspace{5pt}
\label{datasets}
    \begin{tabular}{| p{2cm} | p{3cm} | p{3.5cm} |c|c|c|} \hline
   		{\bf Collection} & {\bf Topics} & {\bf Dataset} & {\bf Runs} &  \!{\bf Judgments}\! & \!{\bf \# Relevant}\! \\ \hline
        
     
        TREC-9 Web Track & 451-500 		& WT10g & 104 &  70070 & 2617 \\ \hline
      
        TREC-2001 Web Track & 501-550 	& WT10g & 97 &  70400 & 3363 \\ \hline
        Robust2003 & 601-650, 50 difficult topics from 301-450 & TREC disks 4\&5 minus Congressional Records &78 & 128796 & 6074 \\ \hline
        Robust2004 & 301-450, 601-700 &  TREC disks 4\&5 minus Congressional Records & 110 & 311410 & 17412 \\ \hline
 	\end{tabular}  
\end{table*}

Robust2003 and Robust2004 collections are particularly well-suited to topic selection experimentation since they have relatively more topics (100 and 249, respectively) than many other TREC collections. 
However, because topics of Robust2003 were repeated in Robust2004, we define a new test collection subset which excludes all Robust2003 topics from Robust2004, referring to this subset as Robust2004$_{149}$.
We use Robust2003 and Robust2004$_{149}$ collections for tuning and testing. When testing on Robust2003, we tune parameters on Robust2004$_{149}$, unless otherwise noted. 
Similarly, when testing on Robust2004$_{149}$, we tune parameters on Robust2003. 
%

\bigskip \noindent 
\textbf{Generation of Training Data}. We generate 100K data records for each topic set size from 0-49 ({\it i.e.}, $N=50$ and $W=100K$ in Algorithm~\ref{algorithm_generate_training_data}) for TREC-9 and TREC-2001 and remove the duplicates. The label range is set to 0-49 ({\it i.e.}, $K = 50$ in Algorithm~\ref{algorithm_generate_training_data}) since each of TREC-9 and TREC-2001 has 50 topics.
We merge the data records generated from each test collection to form our final training data. We use this training data in our experiments unless otherwise stated. 

\bigskip \noindent
\textbf{Parameter Tuning}. To tune parameters of MART, we fix the number of trees to 50 and vary the number of leaves from 2-50 with a step-size of 2. For each of those 25 considered configurations, we build a L2R model and select 50 topics (the standard number of topics in TREC collections) using the tuning set. At each iteration of the topic selection process, we rank the systems based on the topics selected thus far and calculate Kendall's $\tau$ rank correlation vs.\ the ground-truth system ranking. Finally, we selected the parameter configuration which achieves the highest average $\tau$ score while selecting the first 50 topics. 


\bigskip
\noindent
\textbf{Evaluation Metrics.} We adopt MAP@100 and statAP~\cite{pavlu2007practical} in order to measure the effectiveness of IR systems. In computing MAP, we use the full pool of judgments for each selected topic. In computing statAP, the number of sampled documents varies in each experiment and are reported in the corresponding sections. Because statAP is stochastic, we repeat the sampling 20 times and report average results. 

\bigskip
\noindent
\textbf{Baselines}. We compare our approach to two baselines:
\begin{itemize}
\item \textbf{Baseline 1: Random}. For a given topic subset size $M$, we randomly select topics $R$ times and calculate the average Kendall's $\tau$ score achieved over the $R$ trials. We set R to 10K for MAP and 1K for statAP (due to its higher computation cost than MAP). 

\item \textbf{Baseline 2:  \citeauthor{hosseini2012uncertainty}~\citeyear{hosseini2012uncertainty}}. We implemented their method using WEKA library~\cite{hall2009weka} since no implementation is available from the authors. The authors do not specify parameters used in their linear SVM model, so we adopt default parameters of WEKA's sequential minimal optimization implementation for linear SVMs. Due to its stochastic nature, we run it 50 times and report the average performance.

\end{itemize}

In addition to these two baselines, we also compare our approach to 
the \textbf{greedy oracle approach} defined in Section~\ref{greedy_approach} (See  Algorithm~\ref{greedy_algorithm}). This serves as a useful oracle upper-bound, since in practice we would only collect judgments for a topic after it was selected.

\subsection{Selecting A Fixed Number of Topics}\label{select_fix_number_of_topics}

In our first set of experiments, we evaluate our proposed L2R topic selection approach vs.\ baselines in terms of Kendall's $\tau$ rank correlation achieved as a function of number of topics (\textbf{RQ-1}). We assume the full pool of judgments are collected for each selected topic and evaluate with MAP. 

\begin{figure*}[t]
\centering
	\begin{subfigure}{6cm}
		\includegraphics[height=4.5cm, width=6cm]{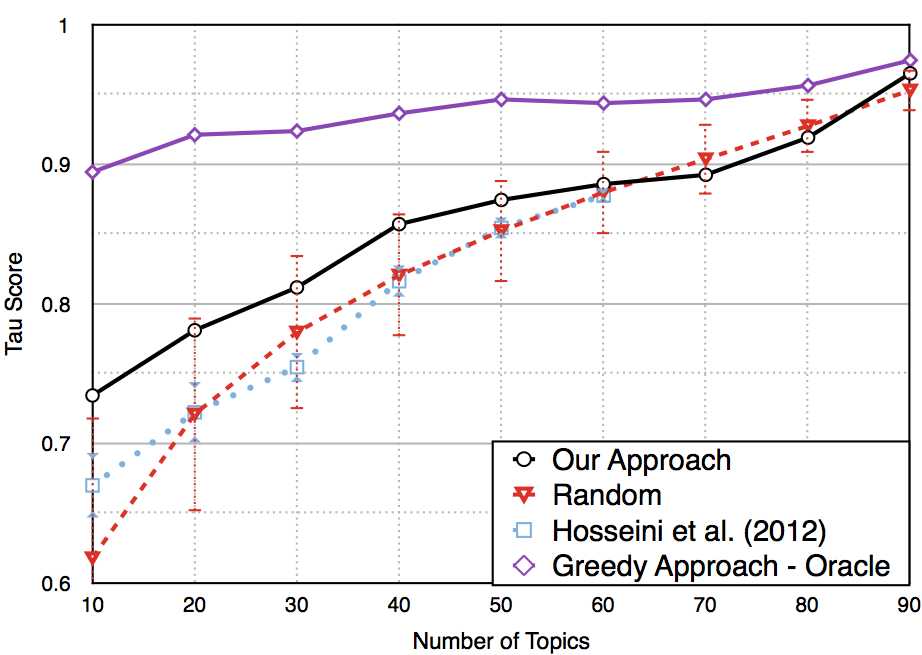} 
  		\caption{Robust2003}
  \end{subfigure}
   \begin{subfigure}{6cm}
		\includegraphics[height=4.5cm, width=6cm]{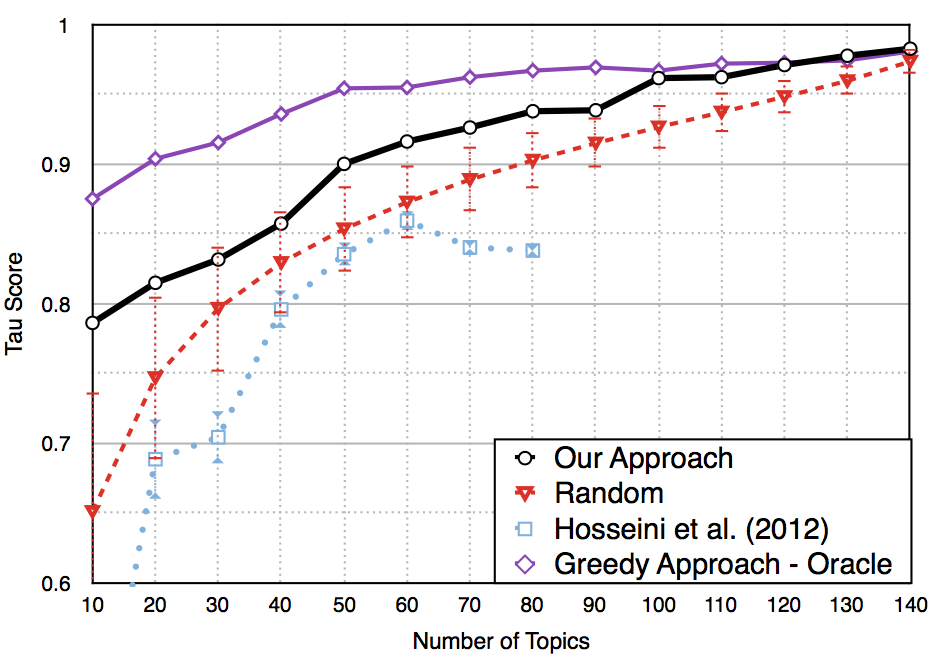} 
  		\caption{Robust2004$_{149}$}
  \end{subfigure}
   \caption{Selecting a fixed number of topics, using full pool judgments per topic, and evaluating with MAP.} 
\label{FIGURE_RQ1_MAP}
\vspace{-10pt}
\end{figure*}

\begin{figure*}[t]
	\centering
 	\begin{subfigure}{6cm}
		\includegraphics[height=4.5cm, width=6cm]{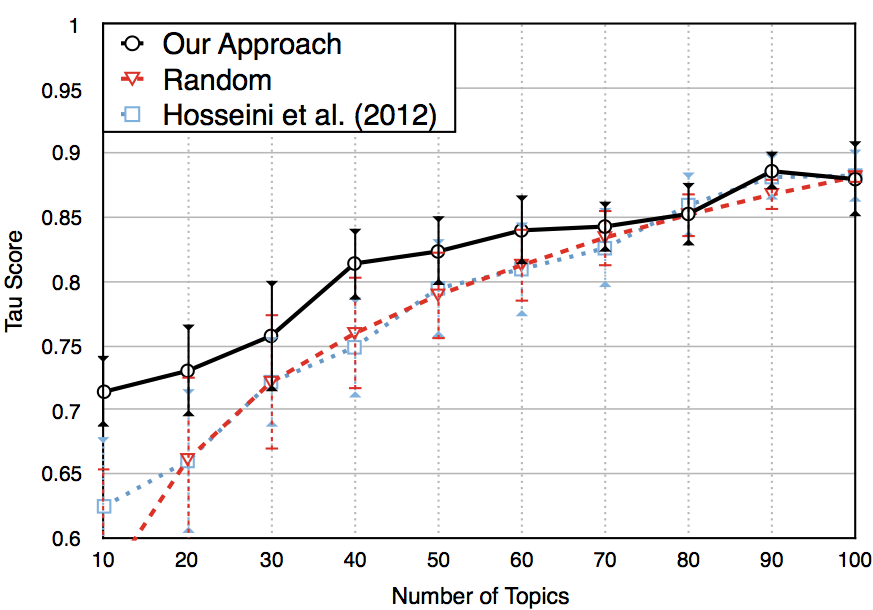} 
  		\caption{Robust2003, 64 Judgments Per Topic}
  	\end{subfigure}
     \begin{subfigure}{6cm}
		\includegraphics[height=4.5cm,width=6cm]{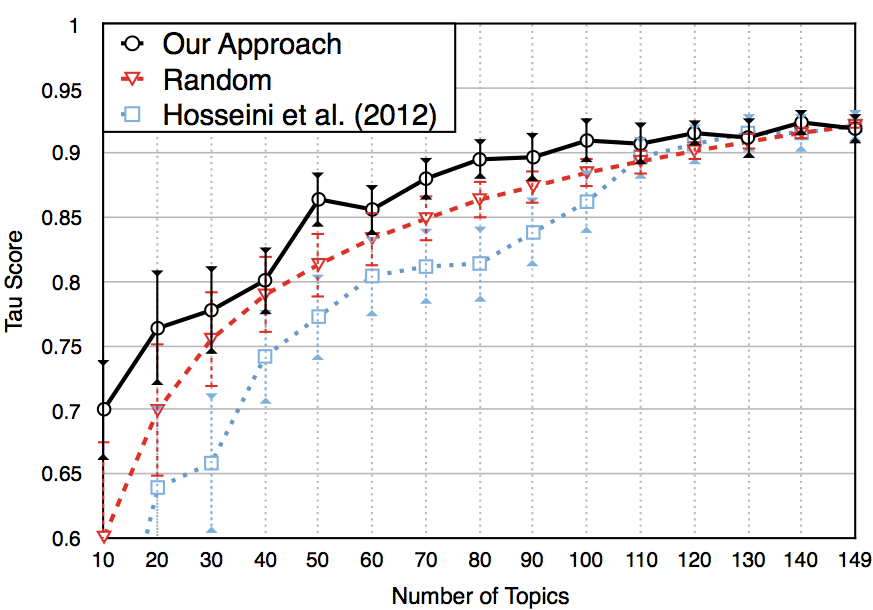} 
 		\caption{Robust2004$_{149}$, 64 Judgments Per Topic}
 	\end{subfigure}
    \begin{subfigure}{6cm}
		\includegraphics[height=4.5cm, width=6cm]{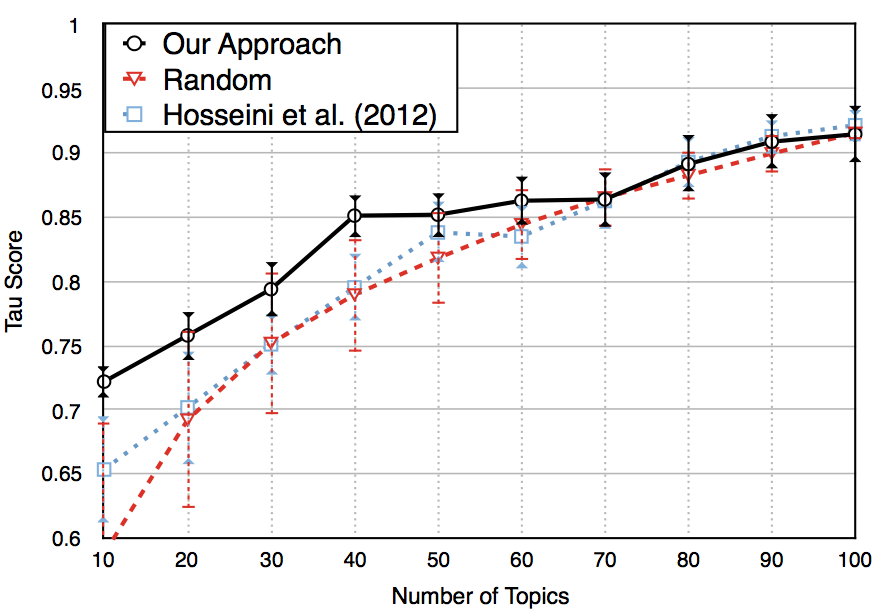} 
  		\caption{Robust2003, 128 Judgments Per Topic}
    \end{subfigure}
    \begin{subfigure}{6cm}
		\includegraphics[height=4.5cm,width=6cm]{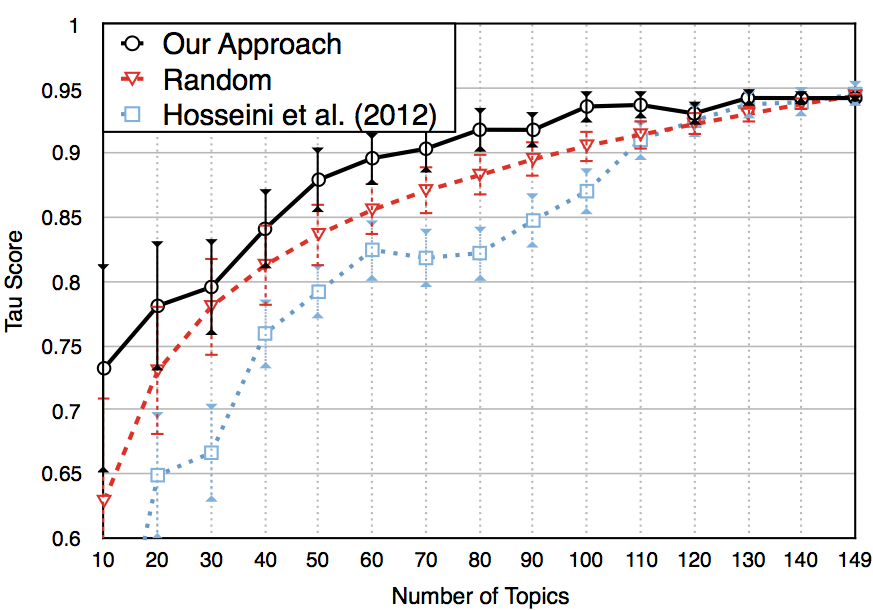} 
 		\caption{Robust2004$_{149}$, 128 Judgments Per Topic}
 	\end{subfigure}
   \caption{Selecting a fixed number of topics and a limited number of documents to judge per topic (using StatAP).}
\vspace{-10pt}
\label{FIGURE_RQ1_STATAP}
\end{figure*}

\textbf{Figure~\ref{FIGURE_RQ1_MAP}} shows results on Robust2003 and Robust2004$_{149}$ collections. Given the computational complexity of \citeauthor{hosseini2012uncertainty}~\citeyear{hosseini2012uncertainty}'s method, which re-trains the classifier at each iteration, we could only select 63 topics for Robust2003 and 77 topics for Robust2004$_{149}$ after 2 days of execution\footnote{Two days is the time limit for executing  programs on the computing cluster we used for experiments.}, so its plots terminate early.
The upper-bound \emph{Greedy Oracle} is seen to achieve 0.90 $\tau$ score (a traditionally-accepted threshold for acceptable correlation~\cite{voorhees2000variations}) with only 12 topics in Robust2003 and 20 topics in Robust2004$_{149}$. 
Our proposed L2R method strictly outperforms baselines for Robust2004$_{149}$ and outperforms baselines for Robust2003 except when 70\% and 80\% of topics are selected. 
Relative improvement over baselines is seen to increase as the number of topics is reduced. This suggests that our L2R method becomes more effective as either fewer topics are used, or as more topics are available to choose between when selecting a fixed number of topics.

 

In our next experiment, instead of assuming the full document pool is judged for each selected topic, we consider a more parsimonious judging condition in which statAP is used to select only 64 or 128 documents to be judged for each selected topic. 
The average $\tau$ scores for each method are shown in \textbf{Figure~\ref{FIGURE_RQ1_STATAP}}. The vertical bars represent the standard deviation across trials. 
Overall, similar to the first set of experiments, our approach outperforms the baselines in almost all cases and becomes more preferable as the number of selected topics decreases. Similar to the previous experiment with full pooling, our L2R approach performs relatively weakest on Robust 2003 when 70 or 80 topics are selected. In this case, our L2R approach is comparable to random selection (with slight increase over it), whereas with the previous experiment we performed slightly worse than random for 70 or 80 topics on this collection. 


We were surprised to see \citeauthor{hosseini2012uncertainty}~\citeyear{hosseini2012uncertainty}'s topic selection method performing worse than random in our experiments, contrary to their reported results. Consequently, we investigated this in great detail. In comparing results of our respective random baselines, we noted that our own random baseline performed $\tau \approx 0.12$  better on average than theirs over the 20 results they report  (using 10, 20, 30, ..., 200 topics), despite our carefully following their reported procedure for implementing the baseline. 
To further investigate this discrepancy in baseline performance, we also ran our random baseline on TREC-8 and compared our results with those reported by \citeauthor{guiver2009few}~\citeyear{guiver2009few}. Our results were quite similar to \citeauthor{guiver2009few}'s. \citeauthor{hosseini2012uncertainty}~\citeyear{hosseini2012uncertainty} kindly discussed the issue with us, and the best explanation we could find was that they took, ``special care when considering runs from the same participant'', so perhaps different preprocessing of participant runs between our two studies may contribute to this empirical discrepancy.

Overall, our approach outperforms the baselines in almost all cases demonstrated over two  test collections. While the baseline methods do not require any existing test collections for training, the existing wealth of test collections produced by TREC and other shared task campaigns make our method feasible. 
Moreover, our experiments show that we can leverage existing test collections in building models that are useful for constructing other test collections. This suggests that there are common characteristics across different test collections that can be leveraged even in other scenarios that are out of the scope of this work, such as the prediction of system rankings in a test collection using other test collections.

\subsection{Feature Ablation Analysis}\label{sec_feature_analysis}

In this experiment, we conduct a feature ablation analysis to study the impact of each core feature and also each group of features on the performance of our approach. 

We divide our feature set into mutually-exclusive subsets in two ways: core-feature-based subsets, and topic-group-based subsets. Each of the core-feature-based subsets consists of all features related to one of our 7 core features (defined in Table~\ref{features_table}). That yields 9 features in each of these subsets; we denote each of them by $\{f\}$, where $f$ represents a core feature. In the other way, we define 5 groups of the topics: the candidate topic $t_c$ (which has 7 core features) and four other groups of topics defined in Section~\ref{features_section} (each has a subset of features using average and standard deviation of the 7 core features, yielding a total of 14 features). We denote each of these feature subsets by $F(g)$, where $g$ represents a group of topics.

In our ablation analysis, we apply leave-one-subset-out method in which we exclude one subset of the features at a time and follow the same experimental procedure with the previous experiments using the remaining features.  We evaluate the effectiveness of systems using MAP.
For each subset of features, we report the average Kendall's $\tau$ correlation over all possible topic set sizes (1 to 100 for Robust2003 and 1 to 149 for Robust2004$_{149}$) to see its effect on the performance. The results are shown in \textbf{Table~\ref{feature_analysis}}.

\begin{table}[htb]
\centering
\caption{Feature ablation analysis. The percentages in parenthesizes show how much the performance is decreased by removing the corresponding subset of features.}
\vspace{5pt}
\label{feature_analysis}
    \begin{tabular}{| l | c | l | l | } \hline
    \multirow{2}{*}{\textbf{Feature Set}} & \bf Number of & \multicolumn{2}{c|}{\bf Average Kendall's $\tau$} \\ \cline{3-4}
   & \bf Features  & Robust2003 &  Robust2004$_{149}$ \\ \hline
     \textbf{All} 		& 63 & 	\bf 0.8535   & \bf  0.9047  \\ \hline 
    All - \{$f_{\bar{w}}$\}	& 54 & 0.8339 (-2\%) & 0.8958 (-1\%)	 \\   
    All - \{$f_{\sigma_w}$\} 	& 54 & 	0.8441 (-1\%)  &	0.8717 (-4\%) \\  
    All - \{$f_{\bar{\tau}}$\}  		& 54 & 	0.8423 (-1\%)  &	0.8796 (-3\%)	\\   
    All - \{$f_{\sigma_\tau}$\} 	& 54	&	0.8465 (-1\%)	 & 	 0.9003 (-0.5\%) \\  
    All - \{$f_\$$\}  			& 54 &	0.8351 (-2\%)	 & 	0.8829 (-2\%)\\  
    All - \{$f_{\sigma_\$}$\} 		& 54	&	0.7929 (-7\%)	 & 	0.8834 (-2\%)\\ 
    All - \{$f_{\sigma_{QPP}}$\} 		& 54 &	0.8118 (-5\%)	 & 	0.8852 (-2\%)\\  
    \hline
    All - $F(t_c)$ & 56			 		&	0.7969 (-7\%)	 & 	0.8485 (-6\%)	 \\ 
    All - $F(P)$ 	& 49				&	0.8310 (-3\%)	 & 	0.8743 (-3\%)	\\ 
    All - $F(\bar{P})$	& 49				&	0.8297 (-3\%)	 & 	0.8882 (-2\%)	\\ 
    All - $F(P\cup \{t_c\})$ 			 		& 49 &	0.7873 (-8\%)	  & 	0.9017 (-0.3\%)	 \\ 
    All - $F(\bar{P}-\{t_c\})$ 					& 49 & 	0.8432 (-1\%)    & 	0.8897 (-2\%)	 \\ \hline
 	\end{tabular}  
\end{table}

The table shows four interesting observations. First, \{$f_{\sigma_\$}$\} and \{$f_{\sigma_w}$\} are the most effective among the core-feature-based subsets, while $F(P\cup \{t_c\})$ and $F(t_c)$ are the most effective among the topic-group-based subsets, when testing on Robust2003 and Robust2004$_{149}$ respectively. Second, \{$f_{\sigma_\tau}$\} has the least impact in both test collections.
Third, the feature subset of the candidate topic $F(P\cup \{t_c\})$ is the best on overage over all subsets, which is expected as it solely focuses on the topic we are considering to add to the currently-selected topics. Finally, testing on both test collections, we achieve the best performance when we use \textit{all} features. 


\subsection{Robustness and Parameter Sensitivity}\label{analysis_of_parameters}
The next set of experiments we report assess our L2R method's effectiveness across different training datasets and parameterizations. We evaluate the effectiveness of systems using MAP.  
In addition to presenting results for all topics, we also compute the average $\tau$ score over 3 equal-sized partitions of the topics. For example, in Robust2004, we calculate the average $\tau$ scores for each of the following partitions: 1-50 (denoted by $\tau_{1-33\%}$), 51-100 (denoted by $\tau_{34-66\%}$) and 101-149 (denoted by $\tau_{67-100\%}$). These results are presented in a table within each figure. 

\begin{figure}
\centering
	\begin{subfigure}{6cm}
		\includegraphics[height=4.5cm, width=6cm]{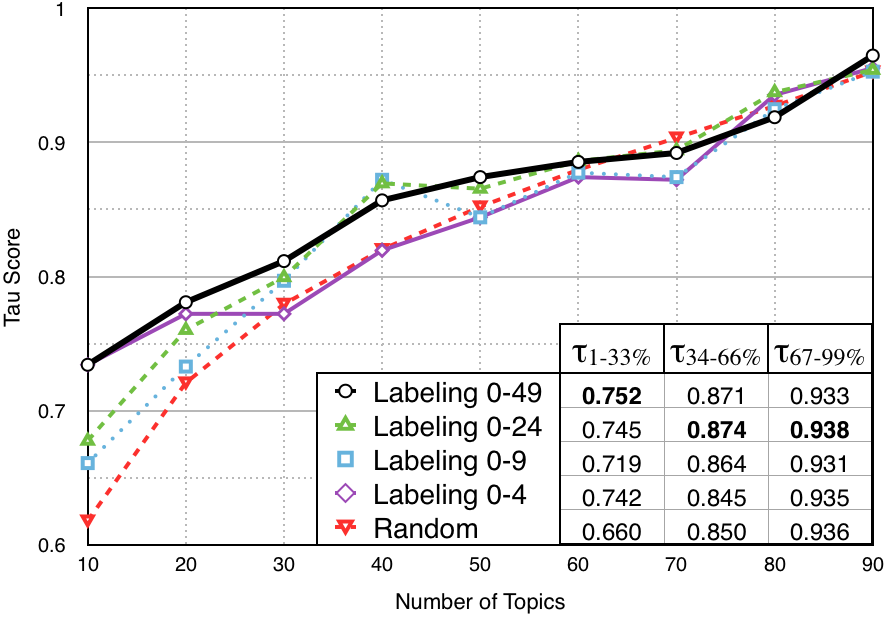} 
  		\caption{Robust2003}
  \end{subfigure}
   \begin{subfigure}{6cm}
		\includegraphics[height=4.5cm, width=6cm]{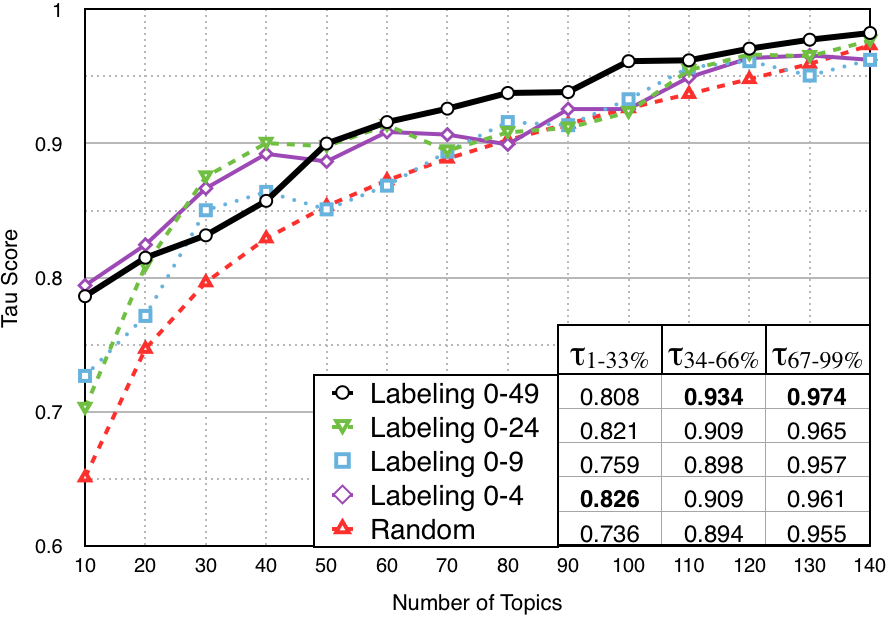} 
  		\caption{Robust2004$_{149}$}
  \end{subfigure}
    \caption{Effect of label ranges in training data for our learning-to-rank model.}.
\label{FIGURE_LABELING}
\vspace{-10pt}
\end{figure}

\textbf{Effect of Label Range in Training Set:} As explained in Section~\ref{generate_training_data}, we can assign labels to data records in various ranges. In this experiment, we vary the label range parameter ($K$ in Line 12 of Algorithm~\ref{algorithm_generate_training_data}) and compare the performance of our approach with the corresponding training data on Robust2003 and Robust2004$_{149}$ test collections. The results are shown in \textbf{Figure~\ref{FIGURE_LABELING}}. It is hard to draw a clear conclusion since each labeling range has varying performances in different cases. For instance, when we use 5 labels only ({\it i.e.}, Labeling 0-4), it has very good performance with few selected topics. As the number of topics increases, its performance becomes very close to the random approach. Considering  the results in Robust2003 and Robust2004$_{149}$ together, using 50 labels ({\it i.e.}, labeling 0-49) gives more consistent and better results than others. Using 25 labels are better than using 10 or 5 labels, in general. Therefore, we observe that \emph{using fine grained labels yields better results with our L2R approach}.

\textbf{Effect of Size of \textit{Tuning} Dataset:} In this experiment, we evaluate how robust our approach is to having fewer topics available for tuning. 
 For this experiment, we randomly select 50 and 75 topics from Robust2003 and remove the not-selected ones from the test collection. We refer to these reduced tuning sets as R3(50) and R3(75). We use these reduced tuning sets for testing on Robust2004$_{149}$. When we test on Robust2003, we perform a similar approach. That is, we randomly select 50, 75, and 100 topics from Robust2004$_{149}$ and follow the same procedure. We repeat this process 5 times and calculate the average $\tau$ score achieved.

\begin{figure}
\centering
	\begin{subfigure}{6cm}
		\includegraphics[height=4.5cm, width=6cm]{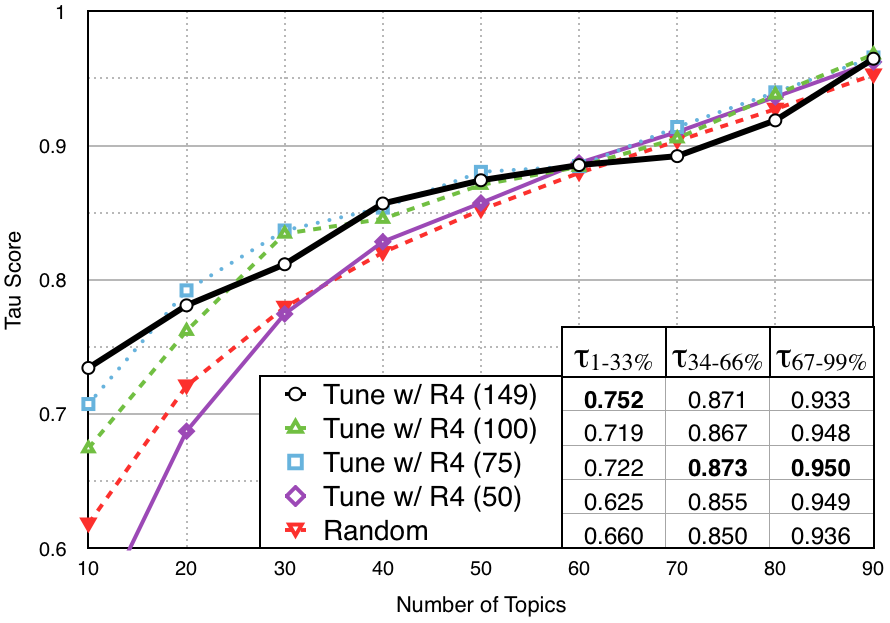} 
  		\caption{Robust2003}
  \end{subfigure}
   \begin{subfigure}{6cm}
		\includegraphics[height=4.5cm, width=6cm]{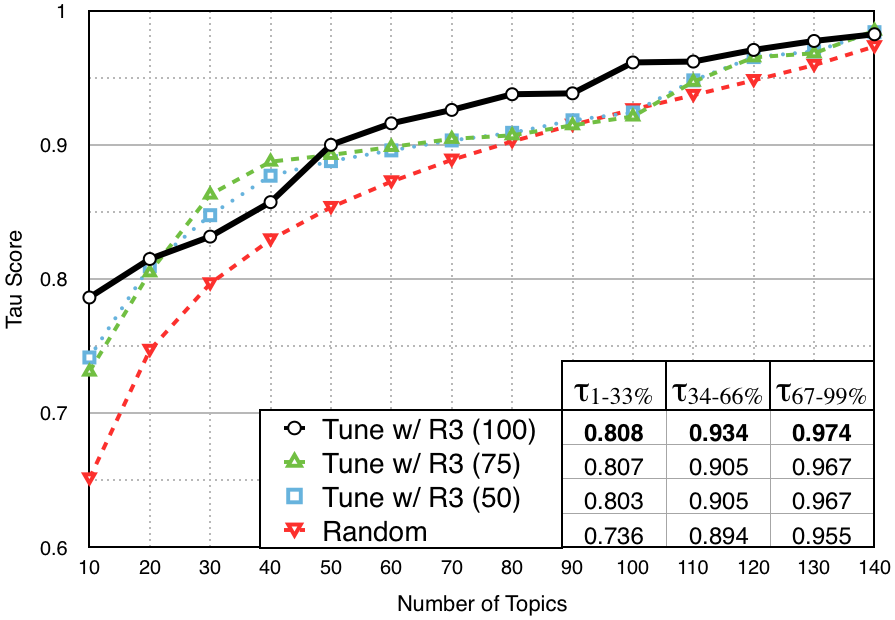} 
  		\caption{Robust2004$_{149}$}
  \end{subfigure}
    \caption{Effect of test collections used for parameter tuning. }.
\label{FIGURE_PARAMETER_TUNING}
\vspace{-10pt}
\end{figure}

The results are presented in \textbf{Figure~\ref{FIGURE_PARAMETER_TUNING}}. The vertical bars represent the standard deviation across 5 different trials. As expected, over Robust2004$_{149}$, we achieve the best performance when we tune with all 100 topics ({\it i.e.}, actual Robust2003); employing 75 topics is slightly better than employing 50 topics. 
Over Robust2003, when the number of selected topics is $\leq$33\% of the whole topic pool size, tuning with 149 topics gives the best results. For the rest of the cases,  tuning with  75 topics gives  slightly better results than  others. As expected, tuning with only 50 topics yields the worst results in general. Intuitively, \emph{using test collections with more tuning topics is seen to yield better results.}



\begin{figure}
\centering
	\begin{subfigure}{6cm}
		\includegraphics[height=4.5cm, width=6cm]{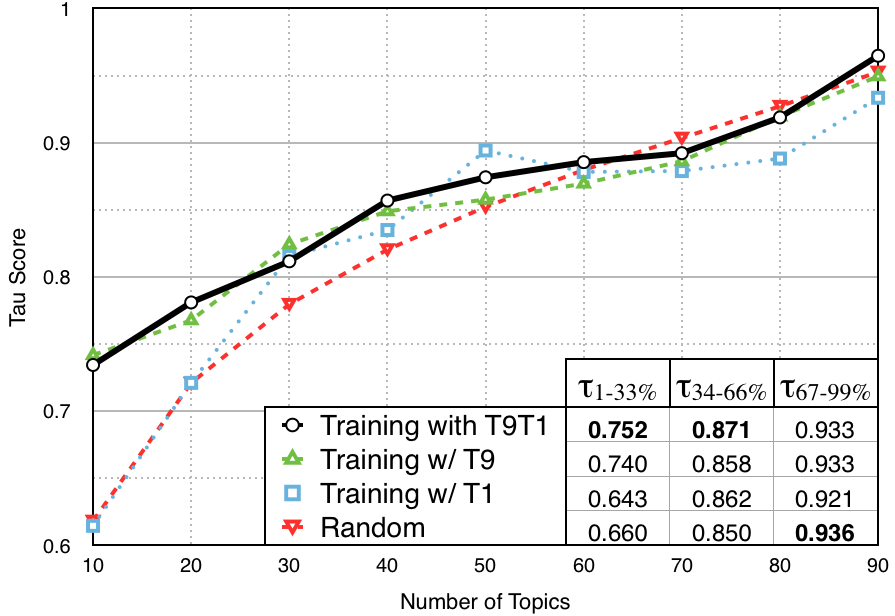} 
  		\caption{Robust2003}
  \end{subfigure}
   \begin{subfigure}{6cm}
		\includegraphics[height=4.5cm, width=6cm]{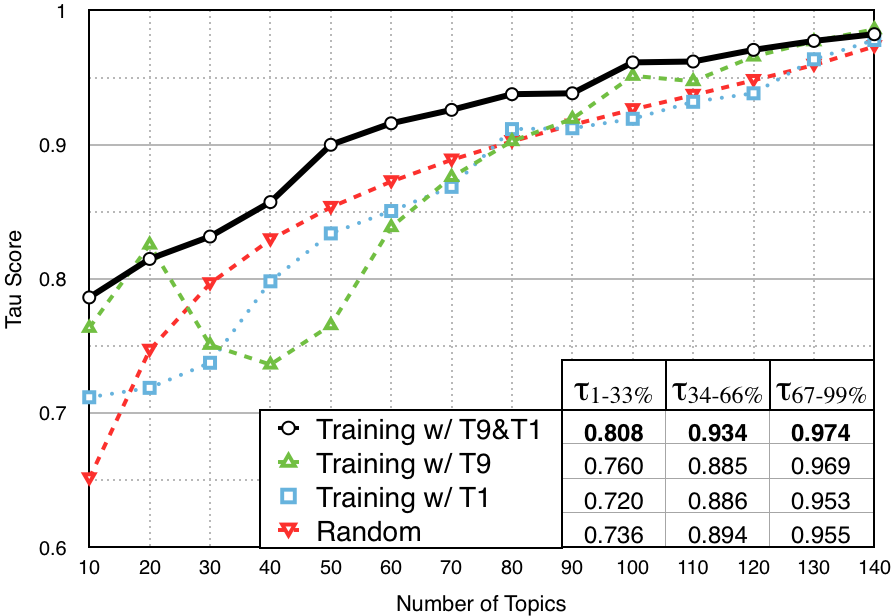} 
  		\caption{Robust 2004$_{149}$}
  \end{subfigure}
    \caption{Effect of test collections used for generating training data.}
    \label{FIGURE_TRAINING_DATA}
\vspace{-10pt}
\end{figure}

\textbf{Effect of Test Collections Used in \textit{Training}:}
In this experiment, we fix the training data set size, but  vary the test collections used for generating the training data. For the experiments so far, we had generated 100K data records for each topic set size from 0-49  with TREC-9 and TREC-2001 and subsequently combined both (yielding 200K records in total). In this experiment, in addition to this training data, we generate 200K data records for each topic set size from 0-49 with TREC-9 and TREC-2001, and use them separately. That is, we have 3 different datasets (namely, T9\&T1, T9 and T1) and each dataset has roughly the same number of data records. The results are shown in \textbf{Figure~\ref{FIGURE_TRAINING_DATA}}. As expected, using more test collections leads to better and more consistent results. Therefore, instead of simply generating more data records from the same test collection, {\it diversifying the test collections in present in the training data appears to increase our L2R method's effectiveness}. 

\subsection{Topic Selection with a Fixed Budget}\label{selecting_with_fixed_budget}

Next, we seek to compare narrow and deep (NaD) vs.\  wide and shallow (WaS) judging when topics are selected intelligently (\textbf{RQ-2}),  considering also familiarization of assessors to topics (\textbf{RQ-3}), the effect of topic generation cost (\textbf{RQ-4}) and judging error (\textbf{RQ-5}).
We evaluate the performance of the methods using statAP~\cite{pavlu2007practical}. 
 The budget is distributed equally among topics. In each experiment, we exhaust the full budget for the selected topics, \textit{i.e.}, as the number of topics increases, the number of judgments per topic decreases, and vice-versa. 

\textbf{Effect of Familiarization to Topics when Judging:}  As discussed in Section~\ref{sec:topic_familiarity}, \citeauthor{carterette2009if}~\citeyear{carterette2009if} found that as the number of judgments per topic increases (when collecting 8, 16, 32, 64 or 128 judgments per topic), the median time to judge each document decreases respectively: 15, 13, 15, 11 and 9 seconds. Because prior work comparing NaD vs.\ WaS judging did not consider variable judging speed as a function of topic depth, we revisit this question, considering how faster judging with greater judging depth per topic may impact the tradeoff between deep vs.\ shallow judging in maximizing evaluation reliability for a given assessment time budget.

\begin{figure}[t]
\centering
	\includegraphics[height=4.5cm,width=6cm]{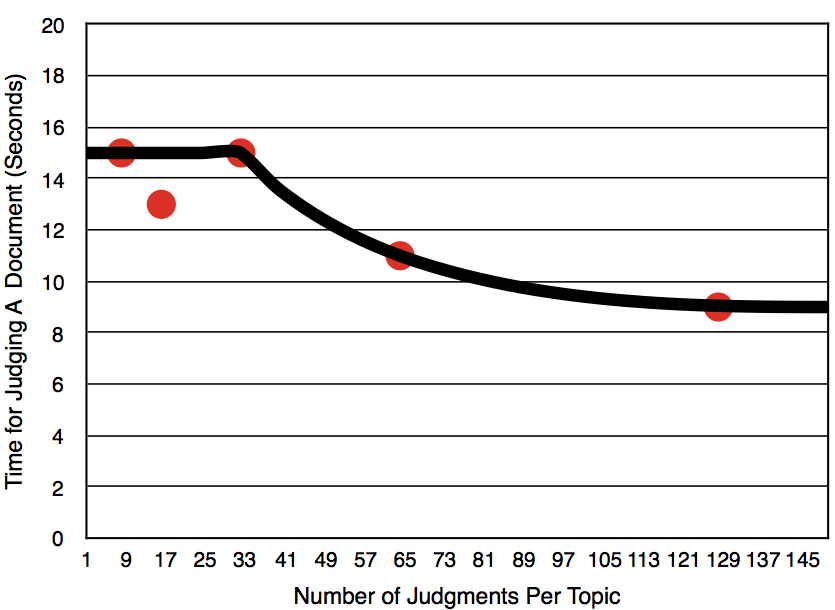} 
 	\caption{Illustration of Equation \protect\ref{equation_judging_speed} 
    modeling the judging time per document as a function of judgments per topic, as fit to actual data points reported by ~\protect\cite{carterette2009if}.}
\label{judging_time}
\vspace{-10pt}
\end{figure}

Using \citeauthor{carterette2009if}~\citeyear{carterette2009if}'s data points, we fit a piece-wise judging speed function ({\bf Equation~\ref{equation_judging_speed}}) to simulate judging speed as a function of judging depth as illustrated in {\bf Figure~\ref{judging_time}}. According to this model, judging a single document takes 15 seconds if there are 32 or fewer judgments per topic ({\it i.e.}, as the assessor ``warms up''). After 32 judgments, the assessors become familiar with the topic and start judging faster. Because judging speed cannot increase forever, we assume that after 128 judgments, judging speed  becomes stable at 9 seconds per judgment.

\begin{equation}
f(x) = 
	\begin{cases}
    	15, & \text{if } x\leq 32\\
        8.761 + 16.856 \times e^{-0.0316 \times x}, & \text{if } 32 < x < 127 \\
        9,  & \text{otherwise}
	\end{cases}
\label{equation_judging_speed}
\end{equation}

For the constant judging case, we set the judging speed to 15 seconds per document. For instance, if our total budget is 100 hours and we have 100 topics, then we spend 1 hour per topic. If judging speed is constant, we  judge $60\ min \times 60 \frac{sec}{min} \div 15 \frac{seconds}{judgment} = 240\ judgments$ for each topic. However, if judging speed increases according to our model, we can judge a larger set of 400 documents per topic in the same one hour.

\begin{figure*}
\centering
	\begin{subfigure}{6cm}
		\includegraphics[height=4.5cm,width=6cm]{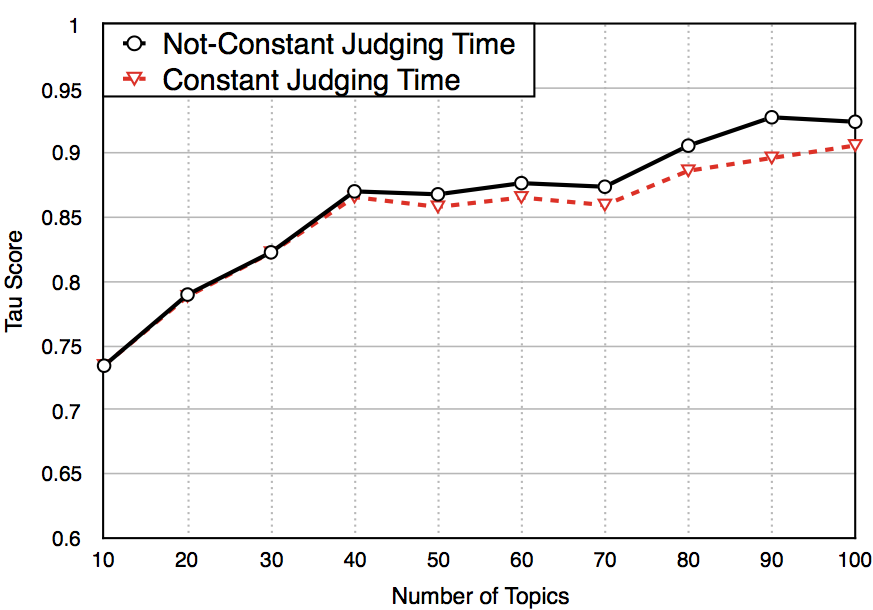} 
  		\caption{Robust2003}
  	\end{subfigure}
   \begin{subfigure}{6cm}
		\includegraphics[height=4.5cm,width=6cm]{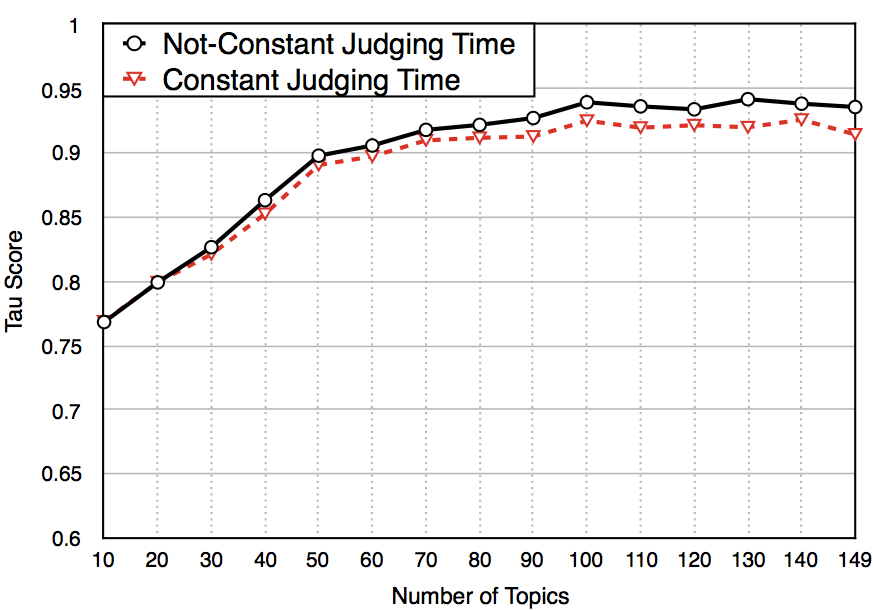} 
  		\caption{Robust 2004$_{149}$}
  	\end{subfigure}
 \caption{Constant vs. Non-Constant Judging Speed.  Topic selection is performed with our method. For each case, we run 20 times and report average  $\tau$ score. }
 \label{FIGURE_CONSTANT_JUDGING}
\vspace{-10pt}
\end{figure*}

We set our budget to 40 hours for both test collections. We initially assume that developing the topics has no cost. 
Results are shown in \textbf{Figure~\ref{FIGURE_CONSTANT_JUDGING}}. We can see that the additional judgments due to faster judging results in a higher $\tau$ score after 30 topics, and its effect increases as the number of topics increases. Since the results are significantly different ($p$ value of paired t-test is 0.0075 and 0.0001 in experiments with Robust2003 and Robust2004$_{149}$, respectively), results suggest that 
{\it topic familiarity (i.e.,\ judging speed) 
impacts optimization of evaluation budget for 
deep vs.\ shallow judging}. 

\textbf{Effect of Topic Development Cost:} As discussed in Section~\ref{related_work}, topic development cost (TDC) can vary significantly. 
In order to understand TDC's effect, we perform intelligent topic selection with our L2R method and vary TDC from 76 seconds (\textit{i.e.}, time needed to convert a query to topic, as reported in ~\cite{carterette2009if}) to 2432 seconds ({\it i.e.},  $32 \times 76$) in geometric order while fixing the budget to 40 hours for both test collections. Note that TREC spends 4 hours to develop a final topic~\cite{ellenvoorheesemail}, which is almost 5 times more than 2432 seconds. 
 We assume that judging speed is constant ({\it i.e.}, 15 seconds per judgment). For instance, if TDC is 76 seconds and we select 50 topics, then we subtract $50 \times 76=3800$ seconds from the total budget and use the remaining for document judging. 

\begin{figure*}[!ht]
\centering
	\begin{subfigure}{6cm}
		\includegraphics[height=4.5cm,width=6cm]{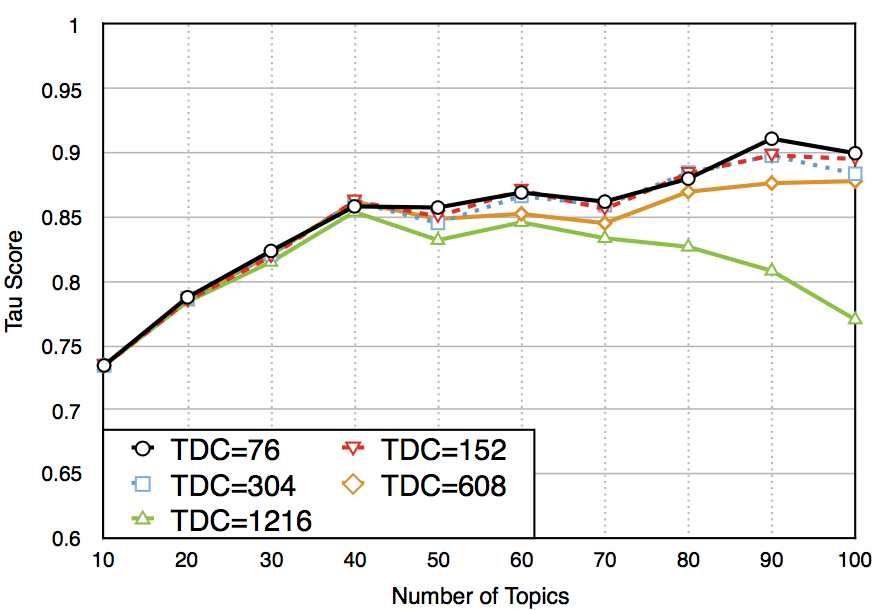} 
 		\caption{Robust 2003}
        \label{TGC_R3_PERFECT_JUDGMENTS}
  	\end{subfigure}
    \begin{subfigure}{6cm}
		\includegraphics[height=4.5cm,width=6cm]{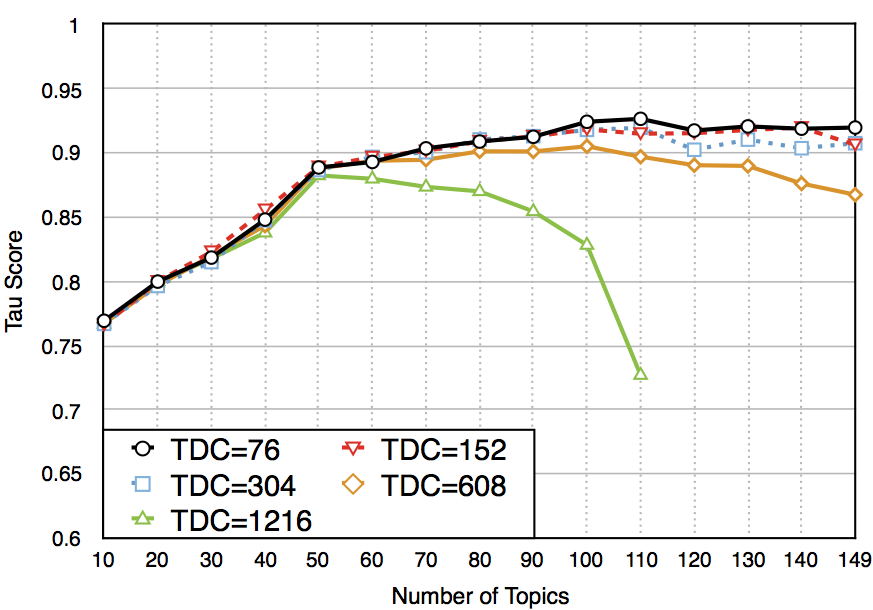} 
  		\caption{Robust 2004$_{149}$}
        \label{TGC_R4_PERFECT_JUDGMENTS}
  	\end{subfigure}
   \caption{Effect of Topic Generation Cost (times listed in seconds). For all cases, we apply statAP and sample the documents 20 times. We assume that each judgment requires 15 seconds. We use our method to select the topics.}
\label{FIGURE_TGC_ANALYSIS}
\end{figure*}

The results are shown in \textbf{Figure~\ref{FIGURE_TGC_ANALYSIS}}. When the topic development cost is  $\leq$152 seconds, results are fairly similar. However, when we spend more time on the topic development, after selecting a number of topics, $\tau$ scores achieved start decreasing due to insufficient budget left for judging the documents. In Robust2004$_{149}$, the total budget is not sufficient to generate more than 118 topics when TDC is 1216 seconds. Therefore, no judgment can be collected when the number of topics is 120 or higher. Considering the results for Robust2004$_{149}$ test collection, when TGC is 304 seconds (which is close to that mentioned in \cite{carterette2008evaluation}), we are able to achieve better performance with 80-110 topics instead of employing all topics. When TDC is 608 seconds, 
using only 50 topics achieves higher $\tau$ score than employing all topics (0.888 vs.\  0.868). Overall, the results suggest that {\it as the topic development cost increases, NaD judging becomes more cost-effective than WaS judging.} 


Another effect of topic development cost can be observed in the \textit{reliability of the judgments}, as discussed in Section~\ref{related_work}. 
In this experiment, we consider a scenario where the assessors rely on the topic definitions and poorly-defined topics can cause inconsistent judgments. In order to simulate this scenario, we assume that 8\% of judgments are inconsistent when $TDC = 76$ seconds. The accuracy of judgments increases by 2\% as TDC doubles. So when TDC = 1216 seconds, assessors can achieve perfect judgments. 
Note that the assessors in this scenario are much more reliable than what is reported in \cite{McDonnell2016}. In order to implement this scenario, we randomly flip over judgments of \emph{qrels} based on the corresponding accuracy of judging. The ground-truth rankings of the systems are based on the original judgments. We set  the total budget to 40 hours and assume that judging a single document takes 15 seconds. We use our method to select the topics. We repeat the process 50 times and report the average. 

\begin{figure*}[htb]
\centering
    \begin{subfigure}{6cm}
		\includegraphics[height=4.5cm,width=6cm]{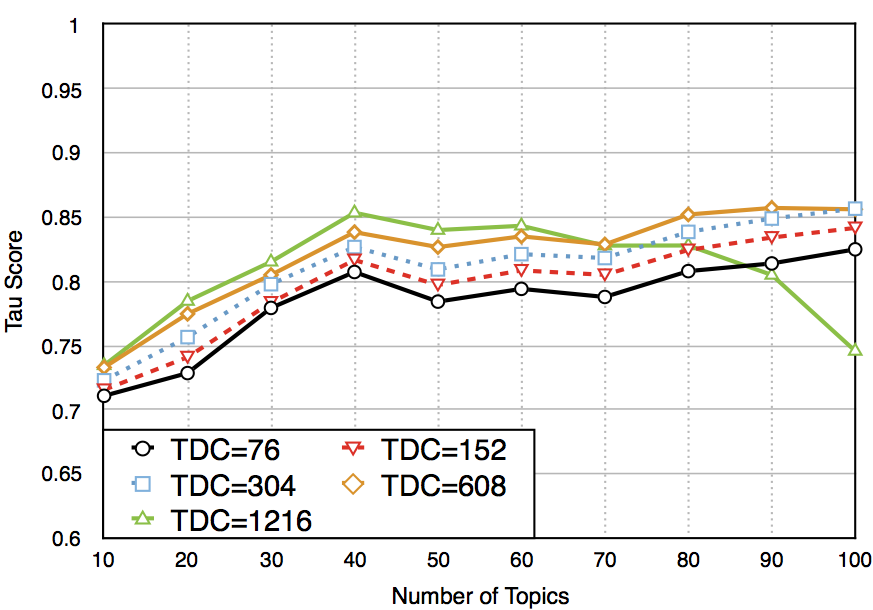} 
  		\caption{Robust 2003}
  	\end{subfigure}
    \begin{subfigure}{6cm}
		\includegraphics[height=4.5cm,width=6cm]{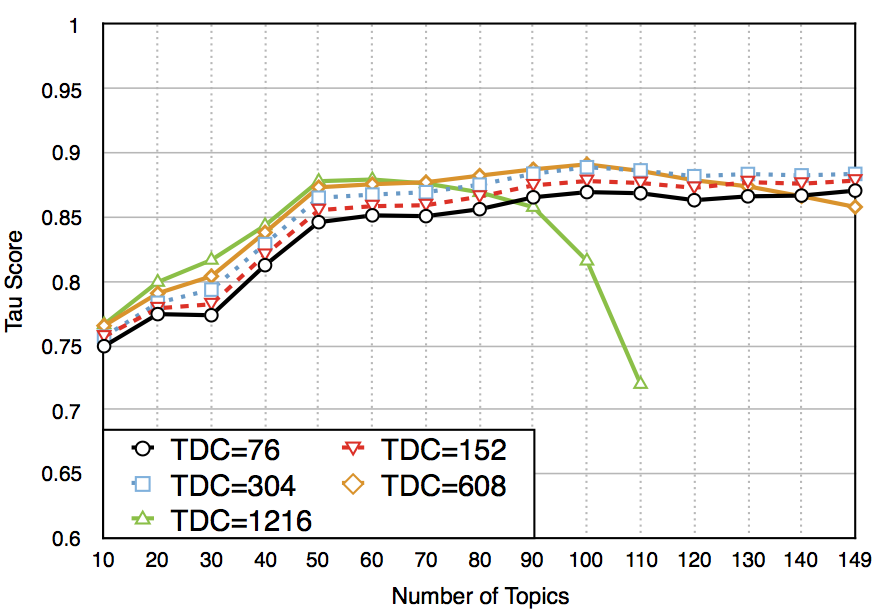} 
  		\caption{Robust 2004 with 149 Topics}
  	\end{subfigure}
   \caption{Effect of topic development cost with imperfect judgments. For all cases, we apply statAP and sample the documents 20 times. We repeat the process 50 times and report the average overall. We assume that each judgment requires 15 seconds. We use our method to select the topics.}
\label{FIGURE_TGC_ANALYSIS_2}
\vspace{-10pt}
\end{figure*}

The results are shown in \textbf{Figure~\ref{FIGURE_TGC_ANALYSIS_2}}. In Robust 2003, the achieved Kendall's $\tau$ scores increase as TDC increases from 76 to 608 seconds due to more consistent judging (opposite of what we observe in Figure~\ref{FIGURE_TGC_ANALYSIS}).
In Robust2004$_{149}$, when the number of topics
is 100 or less, Kendall's $\tau$ score increases as TDC increases from 76 to 608. However, when the number of topics is more than 100, $\tau$ scores achieved with $TDC=608$  start decreasing due to  insufficient amount of budget for judging. We observe a similar pattern in Robust2003 with $TDC=1216$. We achieve the highest $\tau$ scores with $TDC=1216$  and 60 or fewer topics, but the  performance starts decreasing later on as more topics are selected. In general, the results suggest that poorly-defined topics should be avoided if they have negative effect on the consistency of the relevance judgments. However, spending more time to develop high quality topics can significantly increase the cost. Therefore, \emph{NaD becomes preferable over WaS when we target constructing high quality topics}.

\textbf{Varying budget:} In this set of experiments, we vary our total budget from 20 to 40 hours. We assume that the assessors judge faster as they judge more documents, up to a point, based on our model given in Equation~\ref{equation_judging_speed}. We also assume that topic development cost is 76 seconds.

\begin{figure*}
	\centering
   \begin{subfigure}{6cm}
		\includegraphics[height=4.5cm,width=6cm]{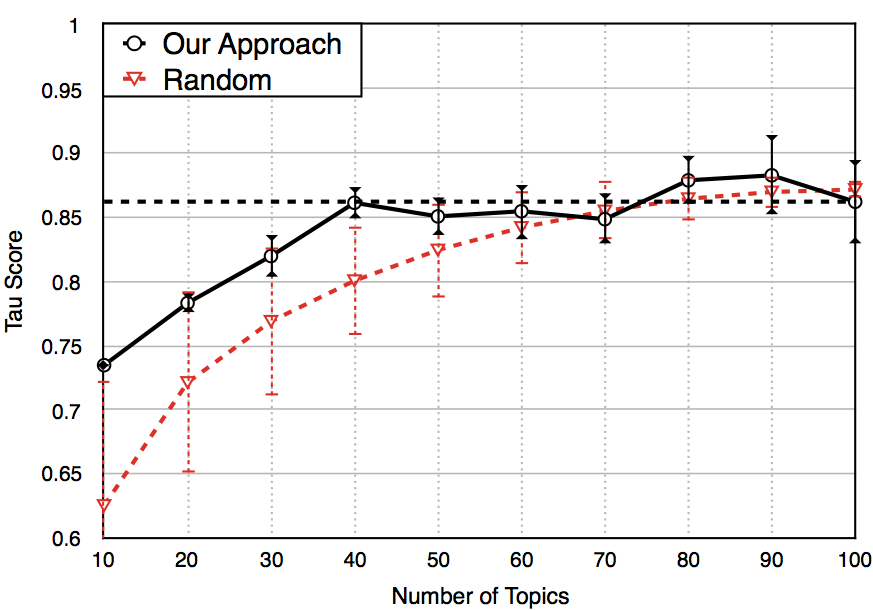} 
  		\caption{Robust 2003 - 20 Hours }
 	\end{subfigure}
  	\begin{subfigure}{6cm}
		\includegraphics[height=4.5cm,width=6cm]{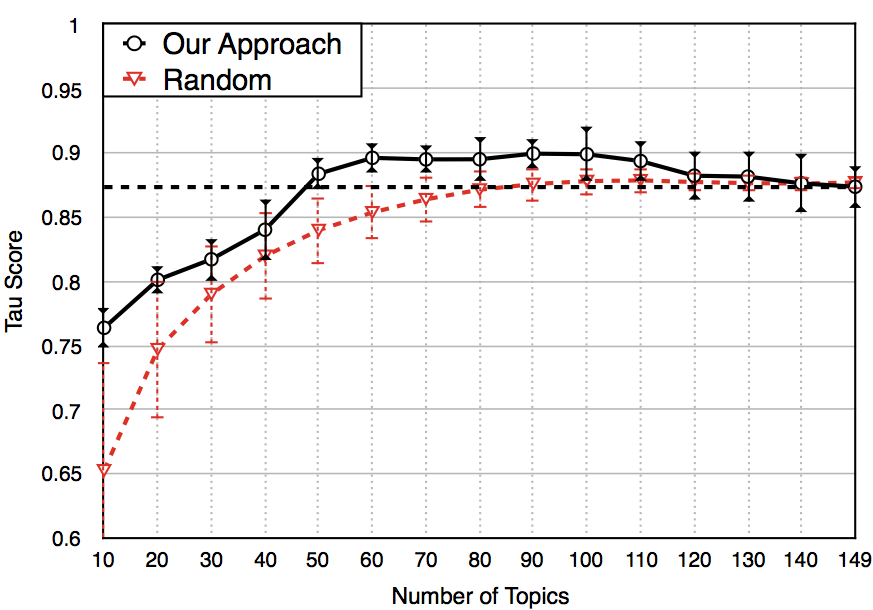} 
  		\caption{Robust 2004$_{149}$ - 20 Hours}
  	\end{subfigure}  
 	\begin{subfigure}{6cm}
		\includegraphics[height=4.5cm, width=6cm]{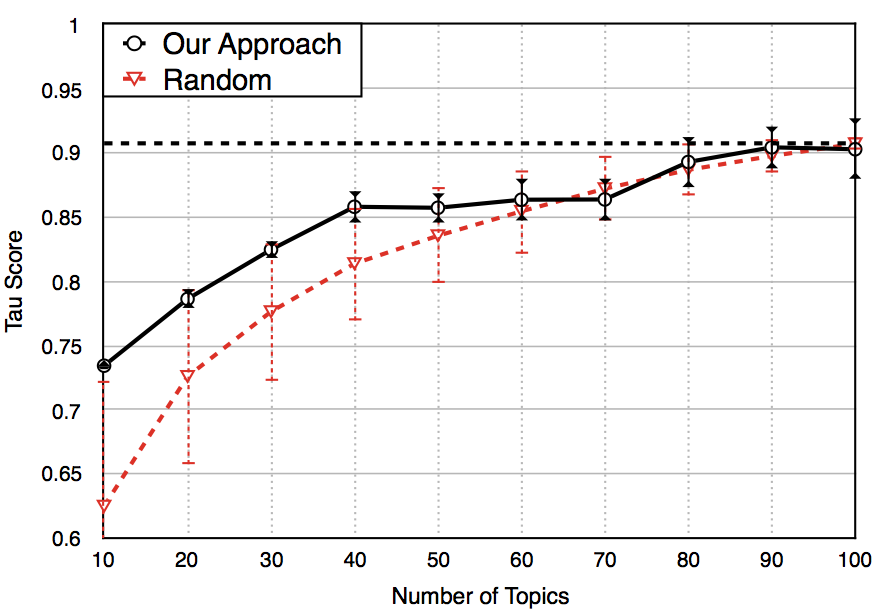} 
  		\caption{Robust 2003 - 30 Hours }
  	\end{subfigure}
     \begin{subfigure}{6cm}
		\includegraphics[height=4.5cm,width=6cm]{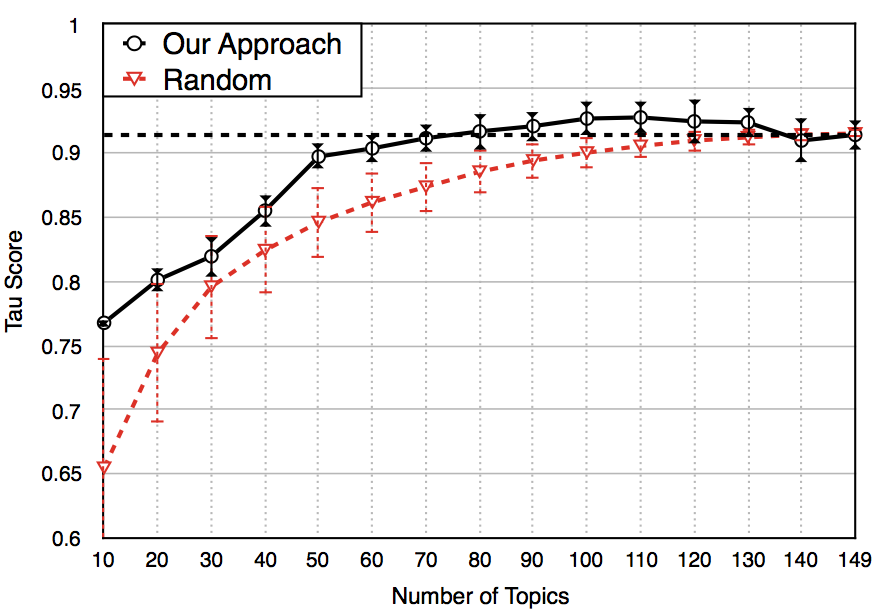} 
 		\caption{Robust 2004$_{149}$ - 30 Hours }
 	\end{subfigure}
    \begin{subfigure}{6cm}
		\includegraphics[height=4.5cm, width=6cm]{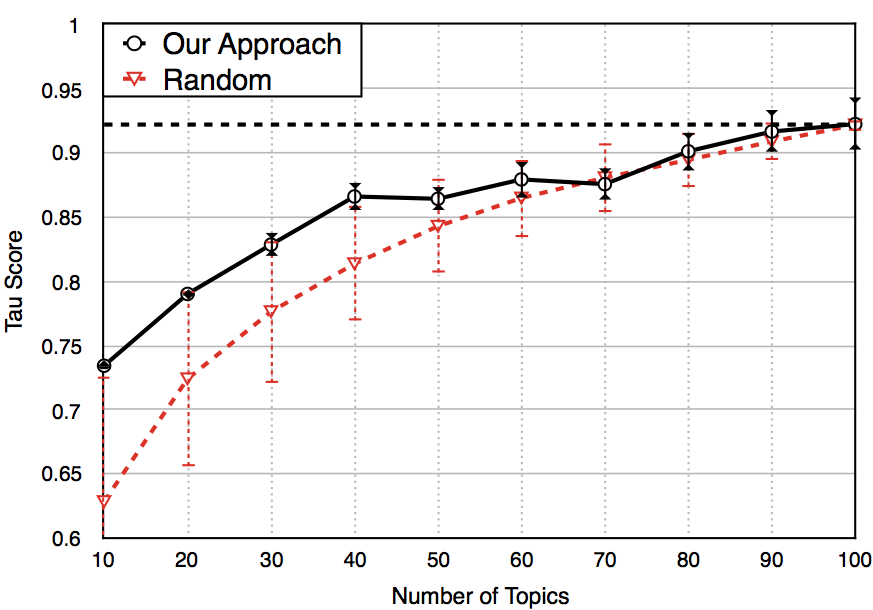} 
  		\caption{Robust 2003 - 40 Hours}
    \end{subfigure}
    \begin{subfigure}{6cm}
		\includegraphics[height=4.5cm,width=6cm]{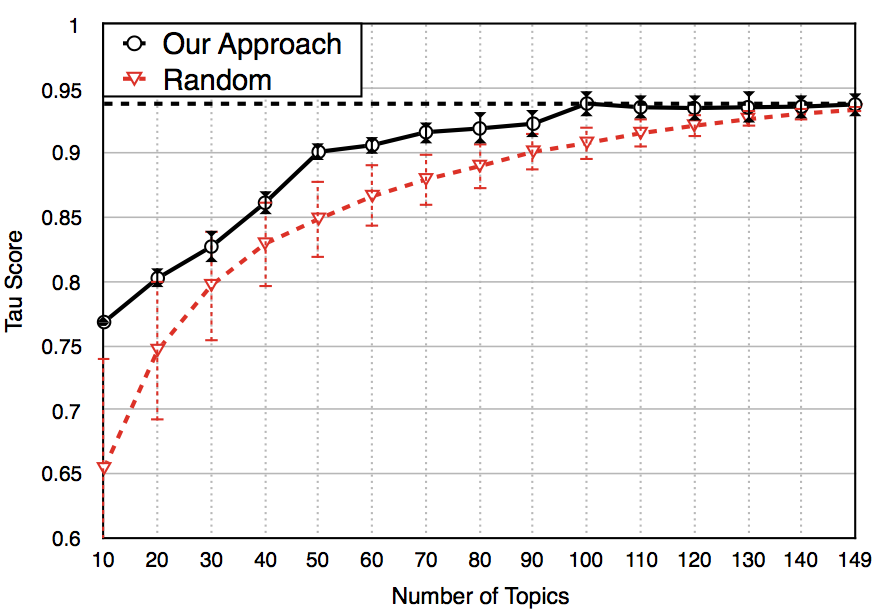} 
 		\caption{Robust 2004$_{149}$ - 40 Hours}
 	\end{subfigure}
   \caption{Topic selection with different test collections when evaluation metric is statAP and we have a fixed budget. For random method, the number of query selection trials is 1000. For both methods, we run statAP 20 times and took the average of all trials. 
  We assume that each judgment requires 15 seconds. The dashed horizontal  line represents the performance when we employed all topics.}
  \label{FIGURE_VARYING_BUDGET}
\vspace{-10pt}
\end{figure*}

The results are shown in \textbf{Figure~\ref{FIGURE_VARYING_BUDGET}}. In Robust2004$_{149}$, our approach performs better than the random method in all cases. In Robust2003, our approach outperforms the random method in the selection of the first 50 topics, while both perform similarly when we select more than 50 topics. Regarding WaS vs.\ NaD judging debate, when our budget is 20 hours, we are able to achieve higher $\tau$ scores  by reducing the number of topics in both test collections. In Robust2004$_{149}$, when our budget is 30 hours, using 90-130 topics leads to higher $\tau$ scores than using all topics. When our budget is 40 hours, we are able to achieve similar $\tau$ scores by reducing the number of topics to 100. 
However, $\tau$ scores achieved by the random method monotonically increase as the number of topics increases (except 20-hours budget scenario with Robust2004$_{149}$ in which using 100-120 topics   achieves very slightly higher $\tau$ scores than using all topics).
That is to say that, {\it WaS judging leads to a better ranking of systems if we select the topics randomly, as reported by other studies~\cite{webber2008statistical,sanderson2005information,bodoff2007test}. However, if we select the topics  intelligently, we can achieve a better ranking by using fewer number of topics for a given budget.} 

 
\textbf{Re-usability of Test Collections:} 
In this experiment, we compare our approach with the random topic selection method in terms of re-usability of the  test collections with the selected topics. We again set topic development cost to 76 seconds  and assume non-constant judging speed. We vary the total budget from 20 hours to 40 hours, as in the previous experiment. 
In order to measure the re-usability of the test collections, we adopt the following process. For each topic selection method, we first select the topics for the given topic subset size. Using only the selected topics, we then apply a leave-one-group-out method~\cite{voorhees2001philosophy}: for each group, we ignore the documents which only that group contributes to the pool and sample  documents based on  remaining documents. Then, the statAP score is calculated for the runs of the corresponding group. After applying this for all groups, we rank the systems based on their statAP scores and calculate Kendall's $\tau$ score compared to the ground-truth ranking of the retrieval systems. We repeat this process 20 times for our method and  5000 times for random method by re-selecting the topics. 

\begin{figure*}
\centering
	\begin{subfigure}{6cm}
		\includegraphics[height=4.5cm,width=6cm]{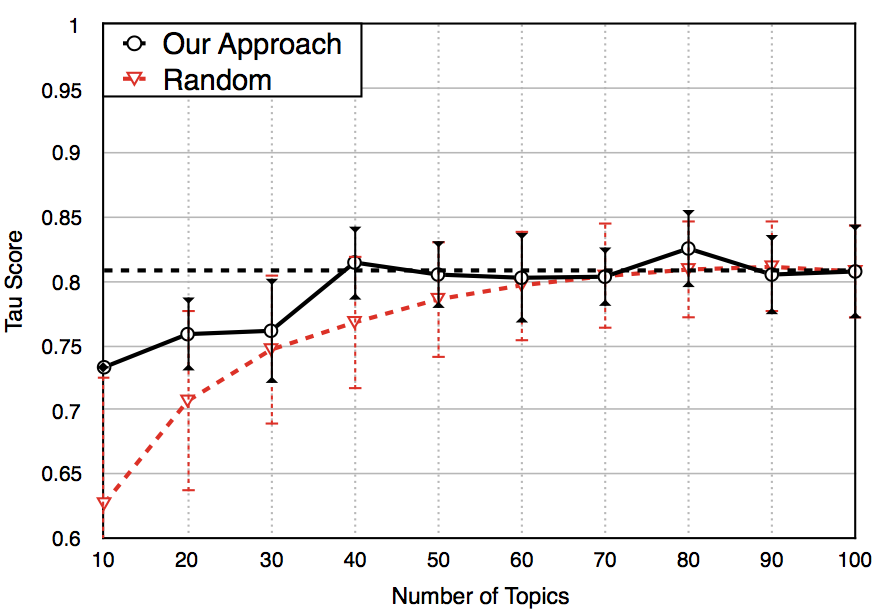} 
  		\caption{Robust 2003 - 20 Hours}
  	\end{subfigure}
	\begin{subfigure}{6cm}
		\includegraphics[height=4.5cm,width=6cm]{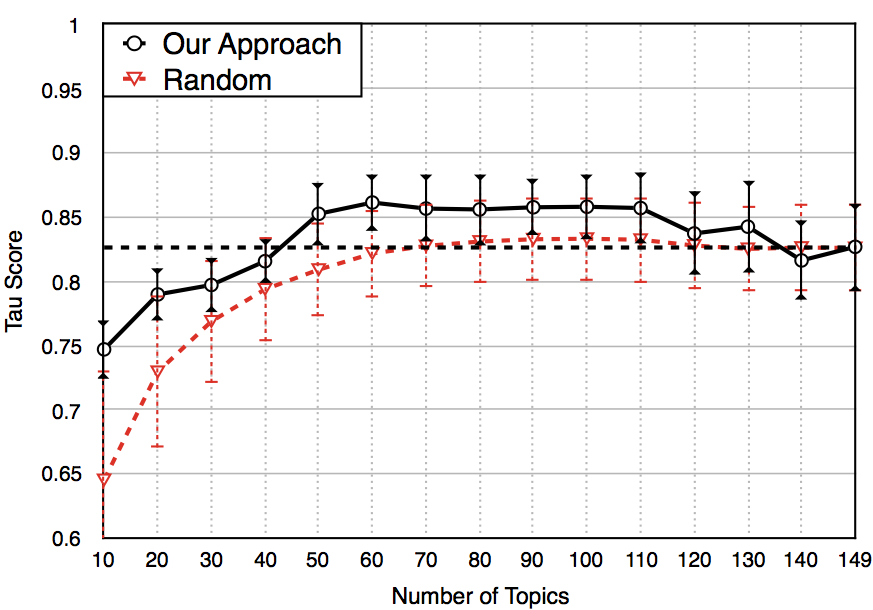} 
  		\caption{Robust 2004$_{149}$  - 20 Hours}
  	\end{subfigure}
    \begin{subfigure}{6cm}
		\includegraphics[height=4.5cm,width=6cm]{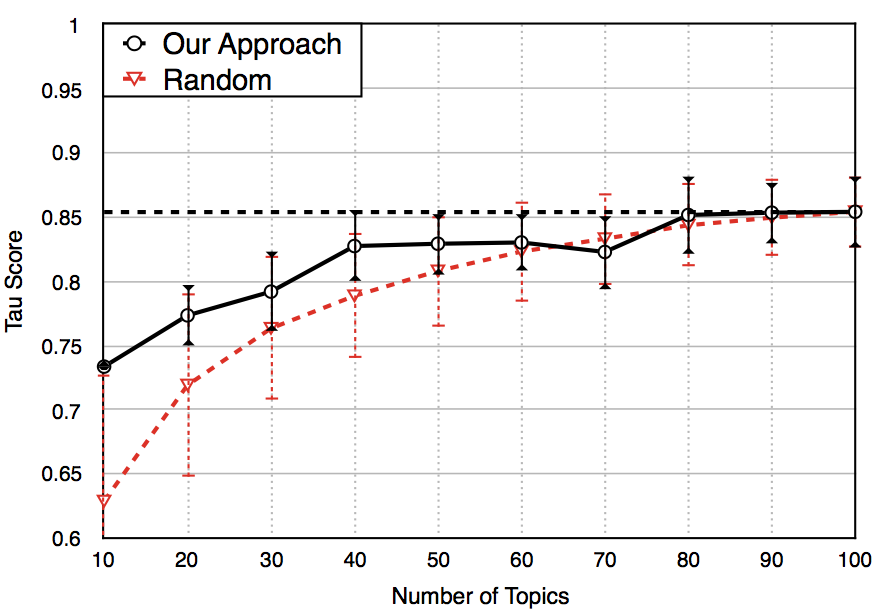} 
 		\caption{Robust 2003 - 30 Hours}
  	\end{subfigure}
    \begin{subfigure}{6cm}
		\includegraphics[height=4.5cm,width=6cm]{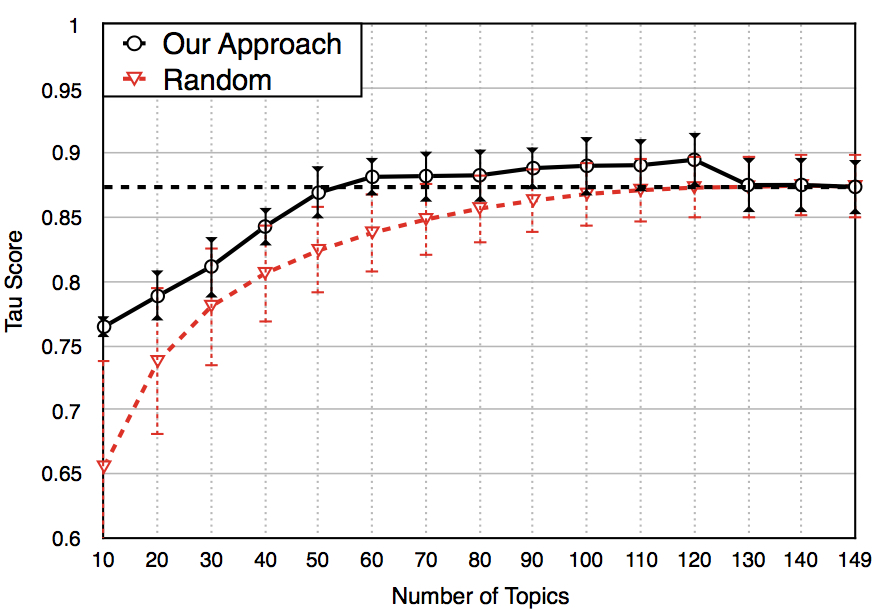} 
 		\caption{Robust 2004$_{149}$ - 30 Hours}
  	\end{subfigure}
    \begin{subfigure}{6cm}
		\includegraphics[height=4.5cm,width=6cm]{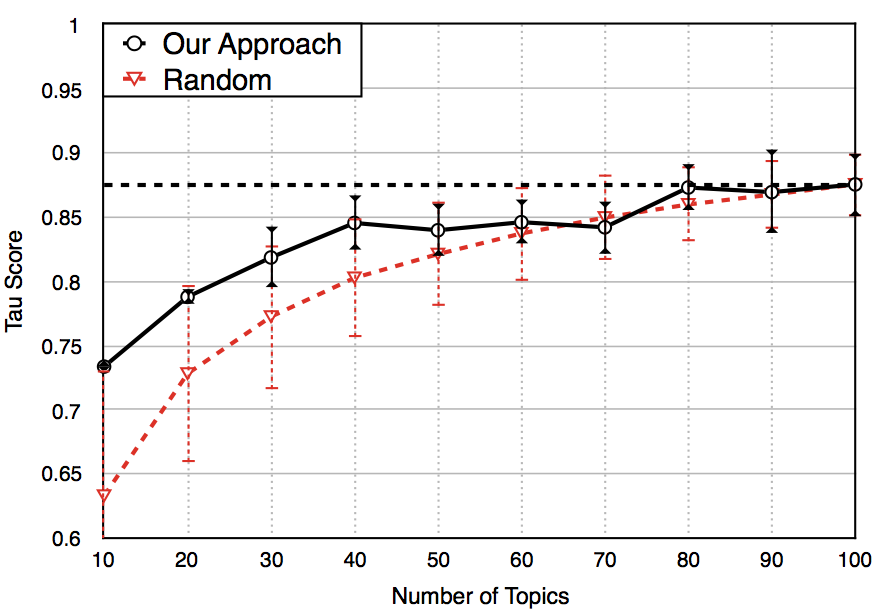} 
  		\caption{Robust 2003 - 40 Hours}
  	\end{subfigure}
    \begin{subfigure}{6cm}
		\includegraphics[height=4.5cm,width=6cm]{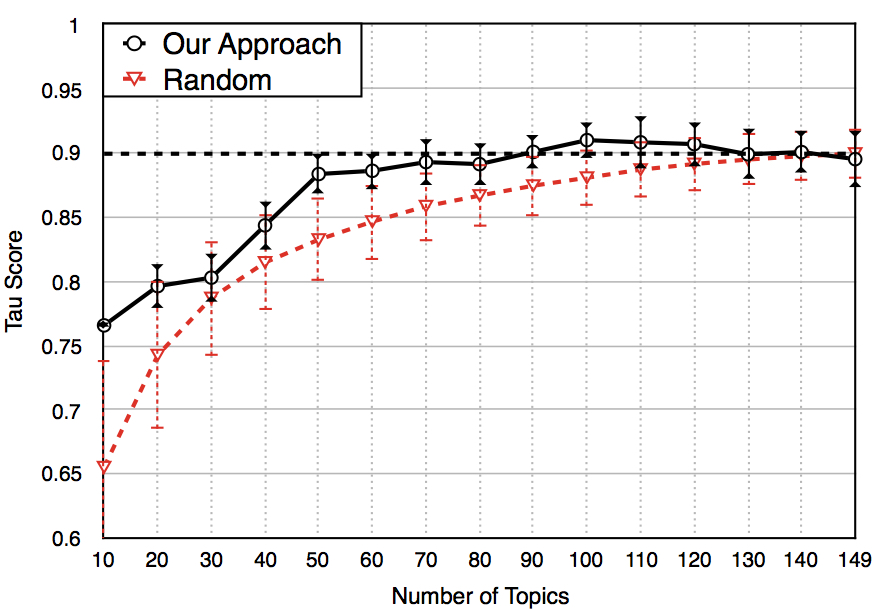} 
  		\caption{Robust 2004$_{149}$ - 40 Hours}
    \end{subfigure}
   \caption{Re-usability Performance. For our method, we run statAP 20 times and take the average of tau scores achieved. For random approach, we select queries 5000 times and calculate statAP once for each query set. Topic generation cost is set to 76 seconds and we assume that the judging speed increases as they judge more documents. The vertical bars represents the standard deviation. The dashed horizontal line represents the performance when we employed all topics.}
\vspace{-10pt}
\label{FIGURE_REUSABLE}
\end{figure*}

The results are shown in \textbf{Figure~\ref{FIGURE_REUSABLE}}. The vertical bars represent the standard deviation and the dashed horizontal line represents the performance when we employ all topics. There are several observations we can make from the results. First,  our proposed method yields more \emph{re-usable}  test collections than random method in almost all cases. As the budget decreases, our approach becomes more effective in order to construct reusable test collections.  
Second, in all budget cases for both test collections, we can reach same/similar re-usability scores with fewer topics.  Lastly,  the $\tau$ scores achieved by the random topic selection method again 
monotonically increases as the number of topics increases in almost all cases. However, by intelligently reducing the number of topics, we can increase the reusability of the test collections in all budget scenarios for Robust2004$_{149}$ and  20-hours-budget scenario for Robust2003. Therefore, \emph{the results suggest that NaD judging can yield more reusable test collections than WaS judging, when topics are selected intelligently.}



\section{Conclusion}\label{conclusion}

While the Cranfield paradigm~\cite{cleverdon1959evaluation} for systems-based IR evaluations has demonstrated remarkably longevity, it has become increasingly infeasible to rely on TREC-style pooling to construct test collections at the scale of today's massive document collections. 
In this work, we proposed a new {\em intelligent topic selection} method which reduces the number of search topics (and thereby costly human relevance judgments) needed for reliable IR evaluation. To rigorously assess our method, we integrated previously disparate lines of research on intelligent topic selection and NaD vs.\ WaS judging. While prior work on intelligent topic selection has never been evaluated against shallow judging baselines, prior work on deep vs.\ shallow judging has largely argued for shallow judging, but assuming random topic selection. Arguing that ultimately one must ask whether it is actually useful to select topics, or should one simply perform shallow judging over many topics, we presented a comprehensive investigation over a set of relevant factors never previously studied together: 1) method of topic selection; 2) the effect of topic familiarity on human judging speed; and 3) how different topic generation processes (requiring varying human effort) impact (i) budget utilization and (ii) the resultant quality of judgments.

Experiments on NIST TREC Robust 2003 and Robust 2004 test collections show that not only can we reliably evaluate IR systems with fewer topics, but also that: 1) when topics are intelligently selected, deep judging is often more cost-effective than shallow judging in  evaluation reliability; and 2) topic familiarity and topic generation costs greatly impact the evaluation cost vs.\ reliability trade-off. Our findings challenge conventional wisdom in showing that deep judging is often preferable to shallow judging when topics are selected intelligently.


More specifically, the main findings from our study are as follows. First, in almost all cases, our proposed approach selects better topics yielding more reliable evaluation than the baselines. Second, shallow judging is preferable than deep judging if topics are selected randomly, confirming findings of prior work. However, when topics are selected intelligently, deep judging often achieves greater evaluation reliability for the same evaluation budget than shallow judging. Third, assuming that judging speed increases as more documents for the same topic are judged, increased judging speed has significant effect on evaluation reliability, suggesting that it should be another parameter to be considered in deep vs.\ shallow judging trade-off. Fourth, as topic generation cost increases, deep judging becomes preferable to shallow judging. 
 Finally, assuming that short topic generation times reduce the quality of topics, and thereby consistency of relevance judgments, it is better to increase quality of topics and collect fewer judgments instead of collecting many judgments with low-quality topics. This also makes deep judging preferable than shallow judging in many cases, due to increased topic generation cost.

As future work, we plan to investigate the effectiveness of our topic selection method using other evaluation metrics, and conduct qualitative analysis to identify underlying factors which could explain {\em why} some topics seem to be better than others in terms of predicting the relative average performance of IR systems. We are inspired here by prior qualitative analysis seeking to understand what makes some topics harder than others \cite{harman2009overview}. Such deeper understanding could provide an invaluable underpinning to guide future design of topic sets and foster transformative insights on how we might achieve even more cost-effective yet reliable IR evaluation.



\section*{Acknowledgments}
This work was made possible by NPRP grant\# NPRP 7-1313-1-245 from the Qatar National Research Fund (a member
of Qatar Foundation). The statements made herein are
solely the responsibility of the authors. We thank the Texas
Advanced Computing Center (TACC) at the University of
Texas at Austin for computing resources enabling this research.

\bibliographystyle{apacite}
\bibliography{References}

\end{document}